\documentclass{aa}
\usepackage{txfonts}
\usepackage[dvips]{graphicx}
\usepackage{rotating}
\newcommand{\asec}{$^{\prime\prime}$}

\def\zctqc{IRAS 05345+3157}
\def\zctst{IRAS 05373+2349}
\def\dcqq{IRAS 18144+3157}
\def\dcu{IRAS 18511+0146}
\def\dcd{IRAS 18517+0437}
\def\dznd{IRAS 19092+0841}
\def\vcvs{IRAS 20126+4104}
\def\vds{IRAS 20216+4107}
\def\vtqt{IRAS 20343+4129}
\def\vdcsd{IRAS 22172+5549}
\def\H{N$_{2}$H$^{+}$}
\def\D{N$_{2}$D$^{+}$}
\def\DCO{DCO$^{+}$}
\def\dens{$n_{\rm H_{2}}$}
\def\AMM{NH$_3$}
\def\ace{CH$_{3}$C$_{2}$H}

\def\CIII{\mbox{C$^{17}$O}}

\def\HII{H{\sc ii}}

\def\kms{\mbox{km~s$^{-1}$}}
\def\cmc{cm$^{-3}$}
\def\cmq{cm$^{-2}$}

\def\solm{\mbox{M$_\odot$}}

\def\Vlsr{$V_{\rm LSR}$}
\def\Dfrac{$D_{\rm frac}$}
\def\Tex{\mbox{$T_{\rm ex}$}}

\begin{document}
%\thesaurus{08(02.13.3; 09.09.1; 09.10.1; 13.19.3)}
\title{Searching for massive pre--stellar cores through observations of \H\ and \D\ 
\thanks{Based on observations carried out with the IRAM Pico Veleta telescope.
IRAM is supported by INSU/CNRS (France), MPG (Germany) and IGN (Spain).}}
%\subtitle{...}
\author{F. Fontani \inst{1} \and  P. Caselli \inst{2} \and
	A. Crapsi \inst{3} 
	\and R. Cesaroni \inst{2}
	\and S. Molinari \inst{4}
	\and L. Testi \inst{2}
	\and J. Brand \inst {1}
        }
\offprints{F. Fontani, \email{fontani@ira.inaf.it}}
\institute{INAF, Istituto di Radioastronomia, CNR, Via Gobetti 101, 
           I-40129 Bologna, Italy \and
	   INAF, Osservatorio Astrofisico di Arcetri, Largo E. Fermi 5,
           I-50125 Firenze, Italy 
	   \and
	   Leiden Observatory, Postbus 9513, 2300 RA Leiden, Netherlands 
	   \and
	   INAF, Istituto di Fisica dello Spazio Interpalenatrio, Via
Fosso del Cavaliere, I-00133 Roma, Italy
%           Infrared Processing and Analysis Center, California Institute
%           of Technology, MS 100-22, Pasadena, CA 91125, USA \and
%           Institut de Radio Astronomie Milimetrique, 300 Rue de la Piscine, 
%           F-38406 St. Martin de Heres, France  \and
%           National Radio Astronomy Observatory, P.O. Box O, Socorro, NM 87801
 }
\date{Received date; accepted date}

%\markboth{Fontani et al.: Mol160}{}
\titlerunning{Massive pre--stellar cores}
\authorrunning{Fontani et al.}

\abstract
{}
{We have measured the deuterium fractionation and the CO depletion factor 
(ratio between expected and observed CO abundance) in a sample of 
high-mass protostellar candidates, in order to understand whether the 
earliest evolutionary stages of high-mass stars have
chemical characteristics similar  to those of low-mass ones.
In fact, it has been found that low-mass starless cores on the 
verge of star formation have large values both of the
column density ratio $N({\rm N_2D^+})/N({\rm N_2H^+})$ and of 
the CO depletion factor.} 
{We have observed with the IRAM-30m telescope and
the JCMT two rotational lines of \H\ and \D ,
the (2--1) line of \CIII\ and \DCO , and the sub-millimeter
continuum towards a sample of 10 high-mass protostellar candidates.}
{We have detected \D\ emission in 7 of the
10 sources of our sample, and found an average value
$N({\rm N_2D^+})/N({\rm N_2H^+})\sim 0.015$. This value is $\sim 3$
orders of magnitude larger than the interstellar D/H ratio, 
indicating the presence of cold and dense gas, in which the 
physical--chemical conditions are similar to those observed in 
low--mass pre--stellar cores. Also, the integrated CO depletion factors
show that in the majority of the sources the expected CO abundances
are larger than the observed values, with a median ratio
of 3.2.}
{In principle, the cold gas that gives origin to the \D\ emission
can be the remnant of the massive molecular core in which the high-mass 
(proto-)star was born, not yet heated up by the central object. 
If so, our results indicate that the chemical
properties of the clouds in which high-mass stars are born are
similar to their low-mass counterparts. Alternatively, this cold gas
can be located into one (or more) starless core (cores) near
the protostellar object. Due to the poor angular resolution of our data,
we cannot decide which is the correct scenario.}


\maketitle

\section{Introduction}
\label{intro}

The initial conditions of the star formation process are still poorly
understood. In recent years, studies of starless low-mass cores
have begun to unveil the physical and chemical features that lead to the 
formation of low-mass stars (Kuiper et al.~\cite{kuiper}; Caselli et 
al.~\cite{caselli99}; Evans et al.~\cite{evans}; Caselli 
et al.~\cite{casellia};~\cite{casellib}; Bergin et al.~\cite{bergin02}; 
Tafalla et al.~\cite{tafalla02};~\cite{tafalla04};~\cite{tafalla06}; 
Harvey et al.~\cite{harvey}; Crapsi et al.~\cite{crapsi}). 
On the other hand, the characterisation
of the earliest stages of the formation of high-mass stars is more 
difficult than for low-mass objects, given their shorter evolutionary 
timescales, larger distances, and strong interaction with their 
environments. 

In the last years various authors (Molinari et al.~2000,~2002; Sridharan et 
al.~2002; Beuther et al.~\cite{beuther}; 
Fontani et al.~2005) have performed extensive studies aimed at the 
identification of precursors of ultracompact (UC)
\HII\ regions, i.e. very young ($<10^{5}$ yr), massive (M$>8$\solm ) 
objects which have not yet ionised the surrounding medium.  
%Molinari et al.~(\cite{mol00}) and 
%Brand et al.~(\cite{brand}), from an initial sample of 260 luminous (i.e.
%with bolometric luminosity higher than $10^{3}$\soll) IRAS 
%sources, found 11 likely precursors of Ultra Compact (UC) \HII\ 
%regions. These sources are characterised by the presence of a compact,
%dense (\dens $>10^{5}$ \cmc) dusty core, seen both in molecular tracers and
%sub-millimeter continuum, and the non-detection of centimeter continuum 
%emission towards the core position.  
In particular, from CS and mm continuum observations, it has been 
noted that in the earliest stages prior to the onset 
of massive star formation, the radial distribution of the intensity 
is quite flat, resembling the structure of starless cores
in more quiescent and less massive molecular clouds (Beuther et 
al.~2002). More evolved 
objects show more centrally-peaked density structures, with power-law 
indices once again resembling those found in low-mass cores (i.e.
\dens$\propto r^{-1.6}$; Shu 1977; Motte \& Andr\'e~2001). 
These results strongly suggest to apply the investigative techniques
that are successful in the study of low-mass star forming cores, 
to the high-mass regime. 

It has been found that when the starless core is on the verge of  
dynamical collapse, most of the high density tracers, including CS
and the CO isotopologues, 
are frozen onto dust grains. CS observations of low-mass cores
clearly show that this molecule is avoiding the central high density 
core nucleus, where the mm continuum peaks (Tafalla et al.~2002).
The morphology of the \CIII\ (1--0) and (2--1) line emission of 
IRAS 23385+6053, a reliable example 
of precursor of an UC \HII\ region (Fontani et al.~2004a), suggests that CO 
is depleted in the high--density nucleus traced by the continuum emission. 
On the other hand, species such as NH$_3$, \H\ and \D\ are good tracers of
the dust continuum emission (Tafalla et al. 2002; Caselli et al. 2002b; Crapsi
et al. 2004, 2005), suggesting that they are not significantly affected by
freeze-out (see Bergin et al.~2001 and Crapsi et al.~2004 for some evidence of
\H\ depletion toward the center of B68 and L1521F, respectively). This is
probably due to the fact that the parent species N$_2$, unlike CO, maintains a
large gas phase abundance at densities around 10$^6$ cm$^{-3}$, despite the
recent laboratory findings that N$_2$ and CO binding energies and sticking
coefficients are quite similar (\"Oberg et al.~2005; Bisschop et al.~2006).
%On the other hand, observations of Nitrogen bearing species in starless 
%low--mass cores (e.g. Tafalla et al.~2002; Caselli et al.~2002b; 
%Crapsi et al.~2004, 2005) have demonstrated that species such as NH$_3$, 
%\H\ and \D\ are not significantly depleted (see Bergin et 
%al.~\cite{bergin01} and Crapsi et al.~2004 for some evidence
%of \H\ depletion toward the center of B68 and L1521F, respectively),
%probably beacuse the parent species N$_2$ is not strongly affected by
%freeze--out unlike CO, despite the recent laboratory findings that
%N$_2$ and CO binding energies and stricking coefficients are quite 
%similar (\"Oberg et al.~\cite{oberg}, Bisschop et al.~\cite{bisschop}).
%In fact, these nitrogen bearing species have been found to be good
%tracers of the dust continuum emission.
The authors above mentioned also found a relation between the column density ratio
D$_{\rm frac}$=N(\D )/N(\H ) and the core evolution: D$_{\rm frac}$ 
is predicted to
increase when the core evolves towards the onset of star formation, 
and then it is expected to drop when the young stellar 
object starts to heat up its surroundings. Summarising, pre--stellar cores 
or cores associated with very young massive stars should show strong emission
in the \H\ and \D\ lines, with large values
of D$_{\rm frac}$, and little or no emission of molecular species such as 
CO and CS, if they evolve like their low-mass counterparts.

Although recently Sridharan et al.~(2005) and Beltr\'an et 
al.~(2006) have detected several candidate high-mass starless 
cores in the neighbourhood of some massive protostar candidate,
a sample of well--identified high-mass pre--stellar cores is lacking.
At present, the objects that are believed to be the closest to
the earliest stages of the high-mass star formation process are those
of two samples of high-mass protostellar candidates,
selected from the IRAS-PSC by Molinari et al.~(\cite{mol96}) and Sridharan 
et al.~(\cite{sridharan}), and then investigated by various authors
(Molinari et al.~\cite{mol98},~\cite{mol00}; Brand 
et al.~\cite{brand}; Fontani et al.~2004a,b; Beuther
et al.~\cite{beuther}; Williams et 
al.~\cite{williams}; Fuller et al.~\cite{fuller}). These studies 
have shown that a large fraction of these sources are 
indeed very young massive objects.
Also, assuming that high-mass stars form in clusters
(Kurtz et al.~\cite{kurtz}), we may expect to find
other cold and dense cores located close to our target sources. 
For this reason, we have observed with the IRAM-30m telescope 
and SCUBA at JCMT a selected sample of candidate precursors of UC \HII\ 
regions in \H , \D , \CIII\ and sub-mm continuum. 
The sources have been selected from the two 
samples mentioned above on the basis of observational features 
indicative of very little evolution: they
are associated with massive ($M\sim 10^{2} M_{\odot}$) 
and compact (diameters less than 0.1 pc) molecular cores, 
in which the kinetic temperatures are around $\sim 20$ K, 
and show faint or no emission in the radio continuum. Also, they have
distances less than 5 kpc.

Sect.~\ref{obs} describes the observations and the data reduction, 
while the results are presented in Sect.~\ref{res} and discussed in 
Sect.~\ref{discu}. A summary of the main findings is 
given in Sect.~\ref{conc}.

\section{Observations and data reduction}
\label{obs}

\subsection{Molecular lines}

All molecular tracers were observed with the IRAM-30m telescope.
In Table~\ref{tab_mol} we give the molecular transitions observed
(Col.~1), the line rest frequencies (Col.~2), the telescope half-power
beam width (HPBW, Col.~3), the channel spacing (Col.~4) and the
total bandwidth (Col.~5) of the spectrometer used.
The observations were made in wobbler--switching mode. Pointing 
was checked every hour. The data were calibrated with the chopper wheel 
technique (see Kutner \& Ulich~\cite{kutner}), with a calibration
uncertainty of $\sim 20\%$.

\subsubsection{\H\ and \D\ }
\label{obsn2h}

A sample of 19 sources, selected from two sample of protostar
candidates as explained in Sect.~1, were 
observed in July 2002, obtaining single-point spectra of the 
\H\ (1--0) and (3--2) lines, and \D\ (2--1) and (3--2) lines towards the 
position of the IRAS source. Then, 
according to the results obtained, in a second observing run carried out
in July 2003, we mapped in on-the-fly 
mode the 10 brightest sources in both the \H\ 
transitions, in order to have a clear identification
of the emission peak position. Finally, spectra of the
\D\ (2--1) and (3--2) lines were obtained with a deep integration 
towards the \H\ (1--0) line peak position. 
These sources are listed in Table~\ref{tab_sou}, while those
observed during the first run only are listed in Table~\ref{tab_sou_2}.

The on-the-fly maps of the \H\ (1--0) and (3--2) transitions were 
obtained in a 1\arcmin\ $\times$1\arcmin\ field centered on
the position listed in Table~\ref{tab_sou}. 
The two transitions were mapped at the same time and registered with
two spectrometers: one with low spectral resolution and large bandwidth
and a second with higher spectral resolution and narrower bandwidth (see
Table~\ref{tab_mol}). The antenna temperature, $T_{\rm A}^{*}$, and the
main beam brightness temperature, $T_{\rm MB}$ are related as
$T_{\rm A}^{*}=T_{\rm MB}\,\eta_{\rm MB}$, with 
$\eta_{\rm MB}=B_{\rm eff}/F_{\rm eff}$=0.80 and 0.48 for the (1--0) 
and (3--2) lines, respectively.
The spectra of the \D\ (2--1) and (3--2) lines were obtained
simultaneously, again with two spectrometers with different spectral
resolutions and bandwidths (see Table~\ref{tab_mol}).
The values of $\eta_{\rm MB}$ are 0.73 and 0.57 for the (2--1) and (3--2) 
transitions, respectively.

The \H\ (1--0), (3--2), and \D\ (2--1), (3--2) rotational transitions 
have hyperfine
structure. To take this into account, we fitted the lines using
METHOD HFS of the CLASS program, which is part of the GAG software
developed at the IRAM and the Observatoire de Grenoble. This method assumes
that all the hyperfine components have the same excitation temperature
and width, and that their separation is fixed to the laboratory value.
The method also provides an estimate of the total optical depth of
the lines, based on the intensity ratio of the different hyperfine
components. The frequency of the \H\ (1--0) line given in 
Table~\ref{tab_mol} is that of the main component (Dore et 
al.~\cite{dore}), while that of the 
(3--2) transition has been taken from Crapsi
et al.~(\cite{crapsi}), and it refers to the $F_{1} F = 4\;5\rightarrow 3\;4$ 
hyperfine component, which has a relative intensity of $17.46\%$.
Rest frequencies of the \D\ lines given in Table~\ref{tab_mol} refer to 
the main hyperfine component (Dore et al.~\cite{dore}).

\subsubsection{\CIII\ and DCO$^{+}$}
\label{obsc17o}

On-the-fly maps of the \CIII\ (2--1) and DCO$^{+}$ (2--1) transitions were
obtained on July 2003. These two transitions and the \H\ (1--0) and (3--2)
lines were mapped at the same time and with two spectrometers. The data were reduced 
using the CLASS program. Like the \H\ lines, the \CIII\ (2--1) line has hyperfine
structure. However, the spectra were too noisy to resolve the different
components, so that the lines were fitted with a single 
Gaussian.

\begin{table*}
\begin{center}
\caption[] {Observed transitions}
\label{tab_mol}
\begin{tabular}{ccccc}
\hline \hline
 molecular  & frequency & HPBW & $\Delta v^{\dagger}$ & Bandwidth$^{\dagger}$ \\
 transition  & (GHz) & (\asec ) & (\kms ) & (\kms ) \\
\hline
\H\ (1$-$0)  & 93.173772 & 26 & $0.031/3.218$ & $113/824$ \\
\H\ (3$-$2)  & 279.511863 & 9 & $0.042/1.073$ & $75/275$ \\
\D\ (2$-$1)  & 154.217137 & 16 & $0.038/1.944$ & $68/498$ \\
\D\ (3--2)  & 231.321966 & 10 & $0.051/1.296$ & $91/332$ \\
\CIII\ (2$-$1) & 224.714385  & 11 & $0.052/1.334$ &  $93/342$\\
DCO$^{+}$ (2$-$1) & 144.077319 & 16 & $0.041/2.081$ & $73/533$ \\
\hline
\end{tabular}
\end{center}
$^{\dagger}$ the two values refer to the two spectrometers used. 
\end{table*}

\subsection{Sub-millimeter continuum}
\label{submm}

Continuum images at 850 $\mu$m of 5 of the 10 sources mapped in
\H\ were taken on 1998 October with SCUBA at the JCMT (Holland
et al.~\cite{holland}). The standard 64--points jiggle map observing
mode was used, with a chopper throw of 2 \arcmin\ in the SE direction.
Telescope focus and pointing were checked using Uranus and the data
were calibrated following standard recipes as in the SCUBA User
Manual (SURF). One of the maps of this dataset, that of \vdcsd ,
has been already published by Fontani et al.~(\cite{fonta2}).

\section{Results}
\label{res}

Parameters and detection statistics of the 10 sources observed
more extensively (see Sect.~\ref{obsn2h})
are listed in Table~\ref{tab_sou}. Column~1 gives the
IRAS name, and the equatorial (J2000) coordinates of the maps
center are listed
in Cols.~2 and 3. In Cols.~4 and 5 we give the source velocities and
kinematic distances, respectively. Cols. 6, 7, 8, 9 and 10 give the following 
information: detection (Y) or non-detection (N) in \H , \D , \CIII ,
DCO$^{+}$ and millimeter or sub-millimeter continuum, respectively. 
In Col.~6 we also give the offset of the position of the \H\ (1--0)
line emission peak with respect to the map center.
As explained in Sect.~\ref{obsn2h}, the \D\ spectra have been taken
toward this position. For the lines mapped,
we have considered as detected those sources showing emission above the 
3$\sigma$ level in the map field. 
Main parameters of the 9 sources observed only in the first observing run are listed
in Table~\ref{tab_sou_2}: the equatorial coordinates in Cols.~2 and 3
represent the observed position, while source velocity, \Vlsr , and 
kinematic distance, $d_{\rm kin}$, are
listed in Cols.~4 and 5, respectively. In Cols. 6 -- 9 we also give the 
peak temperature of the lines detected, or the rms level of the spectrum for 
those undetected.

In the following, we will consider only the results obtained for the
sources listed in Table~\ref{tab_sou}. Since they have been previously 
observed in different tracers 
by other authors, for completeness in Table~\ref{tab_ref} we give 
the most important physical parameters reported in the literature:
the dust temperatures $T_{\rm d}$, obtained from gray-body
fits to the observed spectral energy distribution of the source, 
the gas kinetic temperature $T_{\rm kin}$, derived from \AMM\ observations (Molinari
et al.~\cite{mol96}, Jijina et al.~\cite{jijina}) or \ace\ observations 
(Brand et al.~\cite{brand}),
and H$_2$ column- and volume densities computed from continuum observations.

\begin{table*}
\begin{center}
\caption[] {Source list and detection summary}
\label{tab_sou}
\begin{tabular}{llllllllll}
\hline \hline
IRAS & R.A. (J2000) & Dec. (J2000) & \Vlsr\ & $d_{\rm kin}$ & \H\ $^{\dagger}$ & \D\ & \CIII\ & \DCO\ & (sub-)mm \\
Source & (h m s) & ($o$ $\prime$  $\prime\prime$) & (\kms ) & (kpc) & & & & & \\
\hline
IRAS 05345+3157$^{\bf 1}$ & 05 37 52.4 & 32 00 06 & $-$18.4  & 1.8 & Y (0,$-$5) & Y & Y & Y  & Y \\
IRAS 05373+2349$^{\bf 1}$ & 05 40 24.5 & 23 50 55 & 2.3  & 1.2 & Y ($-$8,0) &     Y & N & Y & Y \\
IRAS 18144$-$1723$^{\bf 1}$ & 18 40 24.5 & $-$17 22 13 & 47.3 & 4.3 & Y ($-$7,0) & N & Y(?) & N & Y \\
IRAS 18511+0146$^{\bf 1}$ & 18 53 38.1 & 01 50 27 & 56.8 & 3.9 & Y (0,$-$5) &      Y$^{\clubsuit}$ & Y & Y & Y \\
IRAS 18517+0437$^{\bf 2}$ & 18 54 13.8 & 04 41 32 & 43.9 & 2.9 & Y (12,0) &        N & Y & N & Y$^{a}$ \\
IRAS 19092+0841$^{\bf 1}$ & 19 11 38.6 & 08 46 30 & 58.0 & 4.5 & Y (8,$-$4) &      N & Y & N & Y \\
IRAS 20126+4104$^{\bf 2}$ & 20 14 26.0 & 41 13 32 & $-$3.8 & 1.7 & Y ($-$5,0) &    Y & Y & Y(?) & Y$^{a,b}$ \\
IRAS 20216+4107$^{\bf 2}$ & 20 23 23.8 & 41 17 40 & $-$2.0 & 1.7 & Y ($-$5,5) &    Y & Y & N & Y$^{a,b}$ \\
IRAS 20343+4129$^{\bf 2}$ & 20 36 07.1 & 41 40 01 & 11.5 & 1.4 & Y ($-8$,4) (13,0) & Y & Y & Y(?) & Y$^{a,b}$ \\
IRAS 22172+5549$^{\bf 1}$ & 22 19 09.0 & 56 04 59 & $-$43.8 & 2.4 & Y ($-$5,0) & Y$^{\clubsuit}$ & Y & N & Y \\ 
\hline
\end{tabular}
\end{center}
\footnotesize
$^{\bf 1}$ = \Vlsr\ and $d_{\rm kin}$ from \AMM\ observations (Molinari et al.~\cite{mol96}) \\
$^{\bf 2}$ = \Vlsr\ and $d_{\rm kin}$ from \AMM\ observations (Sridharan et al.~\cite{sridharan}) \\
$^{\dagger}$ = between brackets, the offset (in arcsec) of the \H\ (1--0) line emission peak 
with respect to the map center (coordinates in Cols.~2 and 3) is given \\
$^{\clubsuit}$ = detected in the \D\ (2--1) transition only \\
(?) = marginal detection \\
$^{a}$ = observed by Beuther et al.~(\cite{beuther}) at 1.2~mm \\
$^{b}$ = observed by Williams et al.~(\cite{williams}) at 850 $\mu$m and/or 450 $\mu$m \\
\normalsize
\end{table*}

\begin{table*}
\begin{center}
\caption[] {Sources observed during the first observing run only. Cols.~6 -- 9
give the peak of each line or the rms level of the spectrum, both in K.}
\label{tab_sou_2}
\begin{tabular}{lllllllll}
\hline \hline
IRAS & R.A. (J2000) & Dec. (J2000) & \Vlsr\ & $d_{\rm kin}$ & \H\ (1--0) & \H\ (3--2) & \D\ (2--1) & \D\ (3--2) \\
Source & (h m s) & ($o$ $\prime$  $\prime\prime$) & (\kms ) & (kpc) & & & \\
\hline
IRAS 05490+2658$^{\bf 2}$ & 05 52 12.9 &  26 59 33 & 0.8 & 2.1 & 0.4 & $\leq 0.9$ & $\leq 0.19$ & $\leq 0.3$ \\
IRAS 18567+0700$^{\bf 1}$ & 18 59 13.6 &  07 04 47 & 29.4 & 2.16 & 0.5 & $\leq 0.4$ & $\leq 0.13$ & $\leq 0.2$\\
IRAS 20106+3545$^{\bf 1}$ & 20 12 31.2 &  35 54 46 & 7.8 & 1.6 & 0.5 & 0.45 & $\leq 0.05$ & $\leq 0.05$ \\
IRAS 20205+3948$^{\bf 2}$ & 20 22 21.9 &  39 58 05 & $-$1.7 & 4.5 & $\leq$0.03 & $\leq 0.1$ & $\leq 0.05$ & $\leq 0.05$ \\
IRAS 20278+3521$^{\bf 1}$ & 20 29 46.9 &  35 31 39 & $-$4.5 & 5.0 & 0.3 & $\leq 0.2$ & $\leq 0.06$ & $\leq 0.08$ \\
IRAS 22134+5834$^{\bf 2}$ & 22 15 09.1 &  58 49 09 & $-$18.3 & 2.6 & 1.2 & 1.6 & $\leq 0.06$ & $\leq 0.08$ \\
IRAS 23139+5939$^{\bf 2}$ & 23 16 09.3 &  59 55 23 & $-$44.7 & 4.8 & 0.9 & 1.5 & $\leq 0.16$ & $\leq 0.3$ \\
IRAS 23151+5912$^{\bf 2}$ & 23 17 21.0 &  59 28 49 & $-$54.4 & 5.7 & $\leq 0.08$ & $\leq 0.5$ & $\leq 0.13$& $\leq 0.36$ \\
%IRAS 23385+6053$^\dagger$ & 23 40 54.498 &  61 10 28.01 & 5.8 & 4.9 & $\leq 0.08$ & $\leq 0.9$ & $\leq 0.19$ & $\leq 0.45$ \\
IRAS 23545+6508$^{\bf 2}$ & 23 57 05.2 &  65 25 11 & $-$18.4 & 0.8 & $\leq 0.07$ & $\leq 0.8$ & $\leq 0.15$ & $\leq 0.42$\\
\hline
\end{tabular}
\end{center}
\footnotesize
$^{\bf 1}$ = \Vlsr\ and $d_{\rm kin}$ from \AMM\ observations (Molinari et al.~\cite{mol96}) \\
$^{\bf 2}$ = \Vlsr\ and $d_{\rm kin}$ from \AMM\ observations (Sridharan et al.~\cite{sridharan}) \\
\normalsize
\end{table*}

\begin{table*}
\caption{Main physical parameters of our sources found in literature:
dust temperature ($T_{\rm d}$), H$_2$ column and volume 
densities ($N_{\rm H_2}$ and $n_{\rm H_2}$, respectively)
and gas kinetic temperature ($T_{\rm kin}$). The
references are shown between brackets.}
\label{tab_ref}
\begin{center}
\begin{tabular}{ccccc}
\hline \hline
Source &  $T_{\rm d}$ & $N_{\rm H_2}$ & $n_{\rm H_2}$ &  $T_{\rm kin}$ \\
       &  (K)   &  ($\times 10^{22}$ \cmq ) & ($\times 10^{5}$ \cmc )   &  (K) \\
\hline
05345+3157 & --  &  8.6 (1) & 1.5 (1) &  17.0 (2) \\
05373+2349 & 27(1)   &   5.9 (1) & 2.4 (1) &  21.2 (3) \\
18144+3157 & 29(1)  &  17.7 (1) & 1.8 (1) & 23.6 (3) \\
18511+0146 & 35 (1) &  6.2 (1) & 0.4 (1) & 27.3 (4) \\
18517+0437 & 38 (5) & 68 (6) & 5.4 (6) & 29 (3) \\
19092+0841 & 29 (1) &  12.4 (1) & 1.6 (1) & 40.4 (4) \\
20126+4104 & 62 (5) &   52 (6) & 11.4 (6) & 23.0 (5) \\
20216+4107 & 46 (5) &   18 (6) &  3.1 (6) & 21.0 (5) \\
20343+4129 & 44 (5) &   22 (6) &  3.9 (6) & 18 (5) \\
22172+5549 & 30 (1) &   4.8 (1) & 0.8 (1) & 17.5 (3) \\
\hline  
\end{tabular}
\end{center}
(1) Molinari et al.~(\cite{mol00})\\
(2) Jijina et al.~(\cite{jijina}) \\
(3) Molinari et al.~(\cite{mol96})\\
(4) Brand et al.~(\cite{brand}) \\
(5) Sridharan et al.~(\cite{sridharan})\\
(6) Beuther et al.~(\cite{beuther})\\
\end{table*}

\subsection{\H\ and \D\ spectra at the \H\ emission peak}
\label{sec_spe}

In the following, we will discuss the data taken with the
spectrometer with the highest spectral resolution.
All sources listed in Table~\ref{tab_sou} have been detected both in the
\H\ (1--0) and (3--2) transitions. Five have also been detected in both
\D\ (2--1) and (3--2) lines, and two (IRAS 18511+0146 and IRAS 22172+5549) 
in the \D\ (2--1) transition only. All spectra of the \H\ (1--0) and 
\D\ (2--1) lines are shown
in the Appendix, in Figs.~\ref{fig_spe1} and \ref{fig_spe2}.

In Table~\ref{tab_lin_n2h} 
we give the \H\ (1--0) and (3--2) line parameters of the spectra taken at the 
peak position of the (1--0) line emission: in \mbox{Cols.~3 -- 7} we list 
integrated 
intensity ($\int T_{\rm MB}{\rm d}v$), peak velocity ($V_{\rm LSR}$), FWHM,
opacity of the main component ($\tau_{10}$), and excitation temperature
(\Tex ) of the \H\ (1--0) line, respectively. \mbox{Cols.~9 -- 12} show 
the same
parameters for the \H\ (3--2) line, for which we do not list \Tex\ 
because we will not use it in the following.
The integrated intensities have been computed over the 
velocity range given in Cols.~2 and 8, while 
for the other parameters we have adopted the fitting 
procedure described in Sect.~\ref{obsn2h}.  
This procedure cannot provide \Tex\ for optically thin lines. However, one
can note that for the sources with opacity $\geq 0.4$,
the values of \Tex\ are comparable to the kinetic temperature 
$T_{\rm kin}$ listed in Col.~5 of Table~\ref{tab_ref},
within a factor $\sim 2$
(with the exception of \dznd , for which the difference is a factor 4).
Therefore, for sources with
optically thin lines, we decided to assume \Tex =$T_{\rm kin}$.

\begin{figure*}
\centerline{\includegraphics[angle=0,width=10cm]{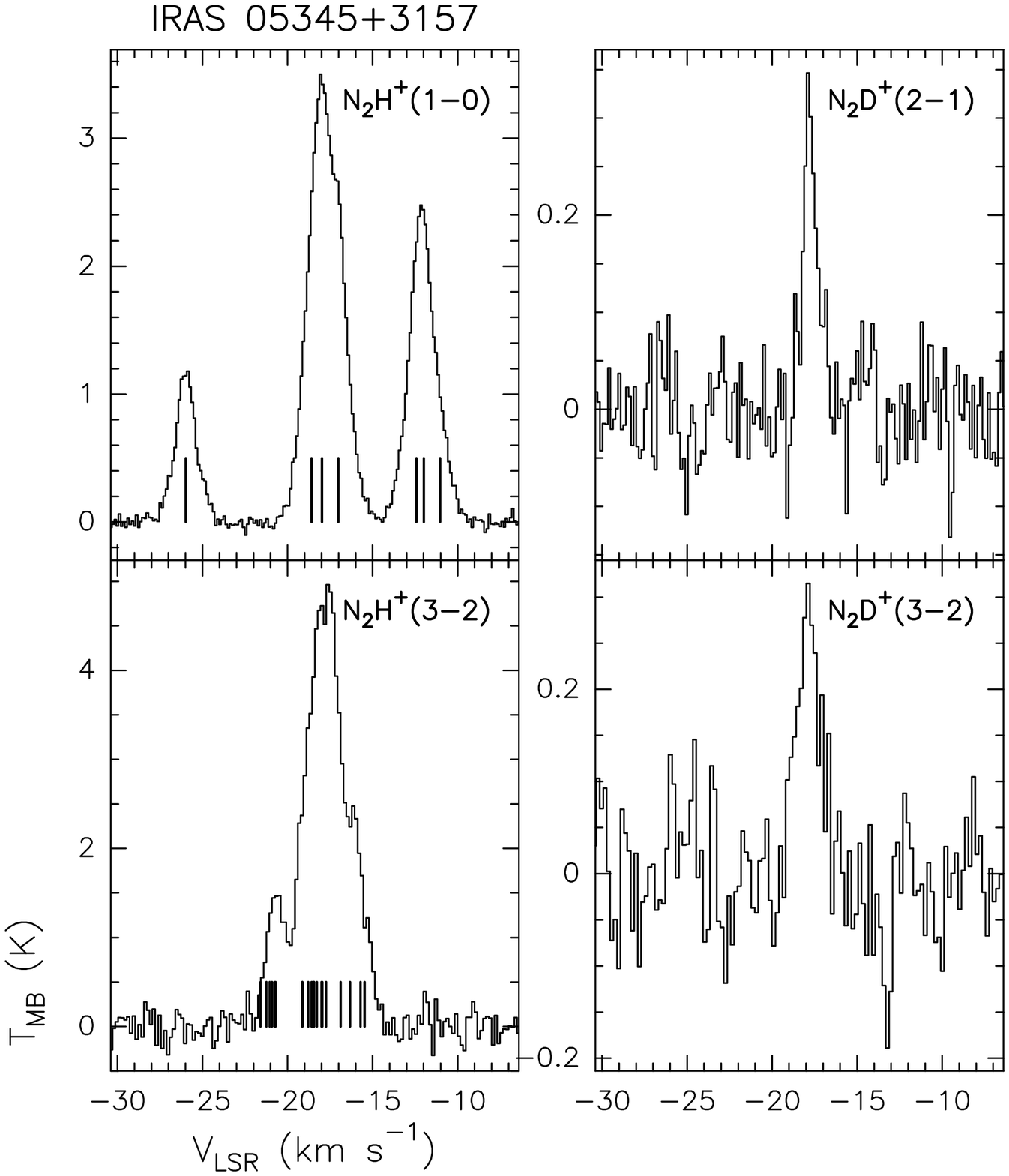}}
\caption{Spectra of IRAS 05345+3157 in \H\ (1--0) and (3--2) lines
(left panel) and \D\ (2--1) and (3--2) lines (right panel). The lines
under the \H\ spectra indicate the position of the hyperfine
components.}
\label{05345_spe}
\end{figure*}

Most of the hyperfine components of the \H\ lines are not well resolved. 
This is due to the fact that the line widths ($\sim 1 - 3$ \kms ) are larger 
than the separation in velocity of most of the different components.
As an example, we show in Fig.~\ref{05345_spe} the spectra of 
IRAS 05345+3157 with the position of the hyperfine components.
Regarding the assumption of equal excitation conditions made to treat the
hyperfine structure of the \H\ (1--0) line, Daniel et al.~(\cite{daniel}) 
have recently found with theoretical calculations that this assumption is 
not valid in 
clouds with densities in the range $10^{4} - 10^{6}$ \cmc\ and 
line optical depths larger than 20. Since all of our sources have 
comparable densities (see Sects.~\ref{sec_dust} and \ref{sec_deut}) 
but the lines have much smaller opacities (see Table~\ref{tab_lin_n2h}), 
the assumption of equal excitation conditions is considered to be valid. 

The line parameters of the \D\ transitions, derived using the
same fitting procedure adopted for \H , are listed in
Table~\ref{tab_lin_n2d}. We give an upper limit for the
integrated intensity of the lines for the undetected
sources, using the observed rms and an average value of the line
FWHM of 1.5 \kms , both for the (2--1) and (3--2) transitions.
With respect to Table~\ref{tab_lin_n2h}, we do not give the line opacities
and \Tex\ because in most of the spectra the uncertainties are 
comparable to or larger
than the values obtained. The two sources with well--determined
opacity of the main component of the \D\ (2--1) line
are \vcvs\ and \vds , for which we obtain
$\tau=0.13\pm 0.09$ and 0.96$\pm 0.03$, respectively.
For the remaining ones, i.e. \zctqc , \zctst , \dcu , IRAS 20343+4139
and \vdcsd , we have fitted
the lines forcing the optical depth to be 0.1.
The observed LSR velocities and line widths are in good agreement with
those derived from \H , indicating that the two molecular species
are tracing the same material.

%\begin{table*}
\begin{sidewaystable*}
\caption{\H\ line parameters}
\label{tab_lin_n2h}
\begin{center}
\begin{tabular}{ccccccccccccc}
\hline \hline
     &  \multicolumn{6}{c}{ \H\ (1--0) } & & & & \H\ (3--2) & & \\
\cline{2-7} \cline{9-13}
source     & vel. range  & $\int T_{\rm MB}{\rm d}v$ & $v_{\rm LSR}$ & FWHM & $\tau_{10}$ & \Tex\ & & vel. range & $\int T_{\rm MB}{\rm d}v$ & $v_{\rm LSR}$ & FWHM & $\tau_{32}$ \\
       & (\kms ) & (K \kms ) & (\kms ) & (\kms ) & & (K) & & (\kms ) & (K \kms ) & (\kms ) & (\kms ) & \\
\hline
05345+3157 & $-$28,$-$9 & 14.66 & $-$17.97 & 1.59 & 0.5$\pm$0.2 & 22.0 & & $-$23,$-$13 & 16.33 & -18.03 & 1.75 & 4.3$\pm$0.5 \\
05373+2349 & $-$8,11 & 20.86 & 2.31 & 1.57 & 0.8$\pm$0.3 & 20.14 & & $-$2,7 & 26.13 & 2.26 & 1.47 & 3.7$\pm$0.2 \\
18144$-$1723 & 36,60 & 38.44 & 48.01 & 3.47 & 0.49$\pm$0.03 & 26.23 & & 42,53 & 31.41 & 48.42 & 3.01 & 7$\pm$1 \\
18511+0146 & 46,68 & 26.76 & 57.12 & 2.86 & 0.1 & -- & & 52,62 & 22.11 & 56.91 & 3.33 & 1$\pm$1 \\
18517+0437 & 33,54 & 26.01 & 43.84 & 2.72 & 0.1 & -- & & 39,48 & 28.38 & 43.71 & 3.21 & 0.7$\pm$0.3 \\
19092+0841 & 45,70 & 13.20 & 57.64 & 3.76 & 0.5$\pm$0.1 & 10 & & 50,63 & 13.13 & 57.07 & 5.84 & 0.1 \\
20126+4104 & $-$14,7 & 36.15 & $-$3.87 & 2.01 & 0.1 & -- & & $-$8,1 & 39.92 & $-$3.36 & 1.57 & 7.82$\pm$0.04 \\
20216+4107 & $-$12,7 & 9.39 & $-$1.66 & 1.30 & 1.2$\pm$0.3 & 9.4 & & $-$6,2 & 14.41 & $-$1.89 & 0.99 & 8.6$\pm$0.5 \\
20343+4139$^{\bf a}$ & 0,22 & 14.22 & 10.87 & 1.43 & 1.79$\pm$0.07 & 9.2 & & 7,14 & 14.45 & 10.81 & 1.21 & 0.1 \\
20343+4139$^{\bf b}$ & 0,22 & 18.39 & 11.48 &  2.44 & 1.37$\pm$0.07 & 9.3 & & 7,14 & 21.51 & 11.75 & 3.69 & 0.1 \\
22172+5549 & $-$54,$-$34 & 6.97 & $-$43.59 & 2.04 & 0.1 & -- & & $-$49,$-$39 & 14.65 & $-$43.56 & 2.63 & 0.1 \\
\hline
\end{tabular}
\end{center}
\footnotesize
$^{\bf a}$ = peak at ($-$8\arcsec ,4\arcsec ) \\
$^{\bf b}$ = peak at (13\arcsec ,0\arcsec ) \\
%\end{table*}
\end{sidewaystable*}
\normalsize

\begin{table*}
\caption{\D\ line parameters}
\label{tab_lin_n2d}
\begin{center}
\begin{tabular}{cccccccccc}
\hline \hline
    & \multicolumn{4}{c}{\D\ (2--1)} & & \multicolumn{4}{c}{\D\ (3--2)} \\
\cline{2-5} \cline{7-10}
  source  & vel. range  & $\int T_{\rm MB}{\rm d}v$ & $v_{\rm LSR}$ & FWHM & & vel. range & $\int T_{\rm MB}{\rm d}v$ & $v_{\rm LSR}$ & FWHM \\
       & (\kms ) & (K \kms ) & (\kms ) & (\kms ) &  & (\kms ) & (K \kms ) & (\kms ) & (\kms ) \\
\hline
05345+3157 & $-$19,$-$16 & 0.33 & $-$17.68 & 1.01 & & $-$20,$-$15 & 0.44 & $-$17.70 & 1.57 \\
05373+2349 & 0,4 & 0.30  & 2.43  & 1.65 & & 0,7 & 0.43 & 2.38 & 1.78 \\
18144$-$1723 &  & $\leq$0.45 & & & & & $\leq$0.68 &  & \\
18511+0146 & 55,60 & 0.23 & 56.90 & 0.99 & & & $\leq$0.54 & & \\
18517+0437 &  & $\leq$0.60 & & & & & $\leq$0.42 & &  \\
19092+0841 &  & $\leq$0.60 & & & & & $\leq$0.68 & &  \\
20126+4104 & $-$6,0 & 0.49 & $-4.14$ & 1.52 & & $-$6,$-$1 & 0.31 & $-$3.86 & 1.05 \\
20216+4107 & $-$4,1 & 0.31 & $-$1.57 & 1.06 & & $-$4,6 & 0.25 & $-1$1.85 & 0.72 \\
20343+4139$^{\bf a}$ & 0,15 & 0.48 & 10.57 & 0.80 & & 9,14 & 0.79 & 10.66 & 1.21 \\
20343+4139$^{\bf b}$ & 8,15 & 0.60 & 11.23 & 2.93 & & 9,15 & 0.84 & 11.62 & 2.17 \\
22172+5549 & $-$45,$-$38 & 0.23 & $-$43.26 & 2.2 & & & $\leq$0.14 & & \\
\hline
\end{tabular}
\end{center}
\footnotesize
$^{\bf a}$ = peak at ($-$8\arcsec ,4\arcsec ) \\
$^{\bf b}$ = peak at (13\arcsec ,0\arcsec ) \\
\end{table*}

\normalsize

\subsection{Distribution of the integrated intensity}

The maps of the integrated intensity of the lines
(\H\ (1--0) and (3--2), \CIII\ (2--1) and \DCO\ (2--1)) are shown 
in the Appendix, from Fig.~\ref{05345_map} to \ref{22172_map}. 
For the sources observed with SCUBA, the emission map of each line
has been superimposed on the continuum 
map at 850$\mu$m, while for the others the position of the 
(sub-)millimeter peaks found in literature are indicated.
We have not shown the maps of the sources marginally detected
in \CIII\ and \DCO\ (2--1) lines.

Figures~\ref{05345_map} - \ref{22172_map} show that in all sources mapped
with SCUBA the distribution of the 850$\mu$m emission and the 
\H\ (1--0) and (3--2) integrated line emission are in good agreement. 
In fact, the emission contours at half of the maximum (FWHM) are fairly well 
overlapping. For almost all of the sources, the angular separation between the 
lines emission peaks and that of the sub-mm emission (or the 1.2~mm for  
\dcd ) is much smaller than the 
beam size, with the exception of the \H\ (3--2) line
in IRAS 18144$-$1723 and IRAS 18517+0437, for which the 
peak of the map is offset by $\sim 12$\asec\ from the dust peak. 
This result confirms that the \H\ molecule and the dust continuum emission
trace similar material, as already found by several authors which 
observed the \H\ lines in both low-mass and high-mass star formation 
regions (Caselli et al.~\cite{casellia}; Crapsi et al.~\cite{crapsi};
Fuller et al.~\cite{fuller}).

The distribution of the \CIII\ (2--1) line integrated 
emission follows that of \H\ and the 850$\mu$m continuum in 
IRAS 18511+0146, IRAS 18517+0437, IRAS 19092+0841, IRAS 20216+4107 and 
IRAS 20343+4129, while in \vdcsd\ it has a different distribution and
in the other sources the signal is faint and irregular. 
Finally, we note that the integrated intensity of the \DCO\ (1--0) line, 
clearly detected only toward IRAS 05345+3157, IRAS 05373+2349 and 
IRAS 18511+0146, presents significant differences from one source to another: 
in IRAS 18511+0146 (Fig.~\ref{18511_map}) the \DCO\ emission is in 
good agreement with that of the
continuum; in IRAS 05345+3157 it is more extended
(see Fig.~\ref{05345_map}), while in IRAS 05373+2349 the two
tracers are almost totally separated (Fig.~\ref{05373_map}). 

Assuming that the emission in each tracer has a Gaussian profile, we have 
derived the angular diameter ($\theta$) of the sources deconvolving
the observed FWHM in that tracer with the appropriate Gaussian beam. The
results are listed in Table~\ref{tab_ang}. We could not compute the
diameters of IRAS 19092+0841 and IRAS 20343+4129 in the \H\ (1--0) line and 
of IRAS 22172+5549 in the \H\ (3--2) line because these are not 
resolved. The diameters derived from the \H\ (1--0) line range from 18.1\asec\
to 41.3\arcsec , while those from the (3--2) transition are between
14.9\asec\ and 29.9\arcsec , indicating that the higher excitation transition
traces a more compact region. The values of $\theta$ derived from \CIII\ 
are typically in between those deduced form the \H\ (1--0) and (3--2) lines,
with the exception of \zctqc\ and \dcd . However, one has to keep in mind
that $\theta$ has been derived under the hypothesis of Gaussian source, which
is only a rough assumption for the \CIII\ emission maps.
For the sources mapped with SCUBA by Williams et al.~(\cite{williams}),
i.e. \vcvs , \vds\ and \vtqt , the authors do not give any estimate
of the angular diameter.  
For this reason, in Table~\ref{tab_ang} we list the diameters
estimated from 1.2~mm continuum maps, derived calculating the geometric 
mean of the major and minor
axes of the sources, obtained by Beuther et al.~(\cite{beuther})
from two-dimensional Gaussian fits (see their Table~2). 
 
Incidentally, we note that the diameters derived from the 
850$\mu$m continuum emission are on average smaller than
those derived from the 1.2~mm continuum by Beuther et al.~(\cite{beuther}).
Since the angular resolutions are comparable,
we believe that this can be due to the different observation technique. 
In fact, as pointed out in 
Sect.~\ref{submm}, the 850~$\mu$m observations have been carried out
in 'jiggle' map observing mode, which provides maps with lower sensitivity 
with respect to the 1.2~mm maps obtained by Beuther et al.~(\cite{beuther})
in the dual-beam on-the-fly observing mode,
and therefore can be less sensitive to extended structures.

\begin{table*}
\caption{Deconvolved source angular diameters (in arcseconds) derived from 
each tracer assuming a Gaussian source.}
\label{tab_ang}
\begin{center}
\small
\begin{tabular}{cccccc}
\hline \hline
source    & \H (1--0) &  \H (3--2) & \CIII\ & (sub-)mm continuum &   \DCO\ \\
\hline
05345+3157 &  31.8 &  20.8 &  7.9 &  21.4  &   34.4 \\
05373+2349 &  34.9 &  21.4 &    -- &  14  &   29.6  \\
18144$-$1723 &  23.8 &  22.2 &   -- &  14.8 &  --  \\
18511+0146 &  41.3 &  29.9 &    41.3 &   29.3 &   23.6 \\
18517+0437 &  25.2 &  18.3 &    28.7 &   29.0$^{\clubsuit}$  &   -- \\
19092+0841  &  --  &  15.1  &   17.6  &  11.7  &  -- \\
20126+4104 &  32.0 &  22.7 &    24.7  &  18.0$^{\clubsuit}$  &   --  \\
20216+4107 &  23.6 &  14.9 &    14.1  & 22.7$^{\clubsuit}$   &   --   \\
20343+4129 &   -- & 21.8 &  32.5  & 26.0$^{\clubsuit}$ &  -- \\
22172+5549 &  18.1 &  -- &    24.0 &  15.7 &   --    \\
\hline
\end{tabular}
\end{center}
${\clubsuit}$ = from 1.2~mm continuum (Beuther et al.~\cite{beuther}). They
are the geometric mean of the major and minor axes obtained from 
two-dimensional Gaussian fits to the observed emission.\\
\vcvs , \vds\ and \vtqt\ have also been observed at 850 $\mu$m by
Williams et al.~(\cite{williams}) but they do not estimate the diameters. \\
\end{table*}

\subsection{Physical properties from dust emission}
\label{sec_dust}

The physical parameters derived form the dust emission are presented in 
Table~\ref{tab_dust}: in Cols.~2 and 3 we give the angular 
and linear diameters ($\theta_{\rm cont}$ and $D$, respectively);
in Cols.~4, 5, 6 integrated flux density ($F_{\nu}$), gas+dust
mass ($M_{\rm cont}$) and visual extinction ($A_{\rm v}$) are listed;
Cols.~7 and 8 give the H$_{2}$ column densities ($N_{\rm H_2}$) derived 
in two different ways which will be explained in the following, and
in cols.~9 and 10 the corresponding H$_2$ volume densities are given. 

The linear diameters have been computed using
the kinematic distances listed in Table~\ref{tab_sou}, and are between $\sim$0.08 and 
$\sim$0.5 pc, typical of clumps hosting high-mass forming stars (Kurtz et al.
~\cite{kurtz}; Fontani et al.~\cite{fonta05}). The flux densities, $F_{\nu}$,
are obtained integrating the SCUBA maps by the
3$\sigma$ level in the maps. For IRAS 20126+4104, IRAS 20216+4107 and
IRAS 20343+4129 we give the values listed in Williams et al.~(\cite{williams}).
%For IRAS 20343+4129, Williams et al.~(\cite{williams}) give a single
%value of the integrated flux density, without separating the two peaks.
%Therefore, the values listed in Table~\ref{tab_dust} are representative of the 
%whole source.
For IRAS 18517+0437, observed neither in this work nor by Williams et 
al.~(\cite{williams}) at 850 $\mu$m, we list the value obtained at 1.2~mm by 
Beuther et al.~(\cite{beuther}).

From the continuum flux density, assuming constant
gas-to-dust ratio, optically thin and isothermal conditions, the total gas+dust mass 
is given by:
\begin{equation}
M_{\rm cont}=\frac{F_{\nu} d^2}{k_{\nu}B_{\nu}(T)}\;.
\label{eqdustmass}
\end{equation}
In Eq.~(\ref{eqdustmass}), $d$ is the distance, $k_{\nu}$ is the dust 
opacity coefficient, derived according to
$k_{\nu}=k_{\nu_0}(\nu/\nu_0)^{\beta}$ (where $\nu_0=230$ GHz and 
$k_{\nu_0}=0.005$ cm$^2$ g$^{-1}$, which implies a gas-to-dust ratio of 
100, Kramer et al.~\cite{kramer}), and
$B_{\nu}(T)$ is the Planck function calculated at the dust temperature 
$T$. For this latter, we decided to use the values listed in 
Col.~5 of Table~\ref{tab_ref} (i.e. the gas temperature derived from \AMM\
or \ace ) instead of the dust temperature given in Col.~2 because this latter has been
derived from the spectral energy distribution of the sources, which
includes the IRAS measurements, not sensitive
to temperatures lower than $\sim 30$ K. Therefore, this value can
overestimate the dust temperature throughout the envelope.
We have assumed $\beta=1.5$, which is a plausible 'average' value
for high-mass star formation regions (see e.g.~Molinari et al.
~\cite{mol00}, Hatchell et al.~\cite{hatchell}, Williams et al.~\cite{williams}).
%For 18517+0437, $T_{\rm d}$ is from Sridharan et al.~(\cite{sridharan})
%and $F_{\nu}$ has been taken from Beuther et al.~(\cite{beuther}).
%For 20126+4104, 20216+4107 and 20343+4129, $T_{\rm d}$ are
%from Sridharan et al.~(\cite{sridharan}) and $F_{\nu}$ from Williams et 
%al.~(\cite{beuther}). For the other sources, $T_{\rm d}$ is from 
%Molinari et al.~(\cite{mol00}).

The visual extinction, $A_{\rm v}$, has been derived from the
continuum flux density using Eq.~(2) of Kramer et 
al.~(\cite{kramer03}): 
\begin{equation}
A_{\rm v}=\frac{1.086}{\kappa^{\prime}}\frac{F_{\rm 850\mu m}}{B_{\rm 850\mu m}(T)\Omega_{\rm s}}\;, 
\label{eq_kramer}
\end{equation}
where $F_{\rm 850\mu m}$ is the integrated flux density
at 850$\mu$m, and $B_{\rm 850\mu m}(T)$ is the Planck function 
calculated at the temperature $T$, for which we have taken the 
gas kinetic temperature (Col.~4 of
Table~\ref{tab_ref}). $\Omega_{\rm s}$ is the source solid angle and 
$\kappa^{\prime}=\kappa_{850}/\kappa_{\rm V}$
is the ratio of absorption and extinction coefficients.
The values of $\Omega_{\rm s}$ have been computed from the angular
diameters given in Table~\ref{tab_ang}, and we
have assumed $\kappa^{\prime}=1.7\times 10^{-5}$, which
is the average value found by Kramer et al.~(\cite{kramer03})
for $\beta=1.5$. For \dcd , for which $F_{\rm 850 \mu m}$ 
is not available, we scaled the flux at 1.2~mm to 850 $\mu$m 
assuming once again a dust opacity index $\beta=1.5$ (i.e. a
spectral index $\alpha=3.5$).  

Finally, we have determined the molecular hydrogen total
column density following two approaches: (1) from $M_{\rm cont}$ and 
$D$ assuming a spherical source; (2) from $A_{\rm v}$ using the relation 
given in Frerking et al.~(\cite{frerking}):
\begin{equation}
N_{\rm H_{2}}=0.94\times 10^{21}\times A_{\rm v}({\rm mag})\;\;\;\;\;\;[{\rm cm^{-2}}]
\end{equation}
We point out that both approaches use $F_{\nu}$ but the latter has the 
advantage of not assuming any gas-to-dust ratio, which is typically affected 
by large uncertainties.

We find masses distributed between $\sim 30$ to $\sim 900$
\solm , visual extinctions from $\sim 90$ to $\sim 10^{3}$ magnitudes,
average column densities between $\sim 10^{23}$ and $\sim 10^{24}$ \cmq\
and average volume densities of $\sim 10^{5}-10^{6}$ \cmc . Mass, visual
extinction and gas
column density were previously estimated from the (sub-)mm continuum 
emission in six sources of our sample by Molinari et 
al.~(\cite{mol00}), and in the remaining ones by Beuther et al.~(\cite{beuther}). 
Even though both Molinari et al. and Beuther et al. used
marginally different assumptions in computing the dust opacity, 
and a different approach to determine $N_{\rm H_2}$ and $A_{\rm v}$, 
their estimates are in good agreement with ours.  
Table~\ref{tab_dust} shows that the column- and volume density estimates derived
from $A_{\rm v}$ is $\sim 3$ times smaller than those obtained from
$M_{\rm cont}$. We believe that this systematic difference is due to the 
different hypotheses made on the dust absorption, since all the other 
parameters used in Eqs.~((\ref{eqdustmass}) and \ref{eq_kramer}) are the same.
However, given the large uncertainties associated with these
measurements, we conclude that the two estimates of $N_{\rm H_2}$ 
are in good agreement, indicating that the assumed
gas-to-dust ratio of 100 is a reasonable value for our sources.

\begin{table*}
\caption{Source properties derived from (sub-)mm continuum emission: deconvolved angular 
diameter ($\theta_{\rm cont}$), integrated 850 $\mu$m continuum flux ($F_{\nu}$, 
for \dcd\ we give the integrated 1.2~mm continuum flux), linear size ($D$), 
gas+dust mass ($M_{\rm cont}$), visual extinction ($A_{\rm v}$) and
H$_{2}$ column and volume densities ($N_{\rm H_2}$ and $n_{\rm H_2}$), 
derived from $M_{\rm cont}$ and $A_{\rm v}$.}
\label{tab_dust}
\begin{center}
\small
\begin{tabular}{cccccccccc}
\hline \hline
source & $\theta_{\rm cont}$ & $D$ & $F_{\nu}$ & $M_{\rm cont}$ & $A_{\rm v}$  & $N_{\rm H_2}^{\dagger}$ & $N_{\rm H_2}^{\dagger\dagger}$ & 
$n_{\rm H_2}^{\dagger}$ & $n_{\rm H_2}^{\dagger\dagger}$ \\
       & (arcsec) & (pc) & (Jy) & (\solm )  &  (mag) & ($\times 10^{23}$\cmq ) & ($\times 10^{23}$\cmq ) & ($\times 10^{5}$\cmc ) & ($\times 10^{5}$\cmc ) \\
\hline
05345+3157  &	   21.4  & 0.19 &  4.42   &  179  &  220 &  6.26 & 2.07 & 10.9 & 3.6 \\
05373+2349  &	     14  & 0.08 &  2.87  &    35 &  239  &  6.81 & 2.25 & 27.8 & 9.3 \\
18144$-$1723  &	   14.8  &  0.31 &  7.76  &    1118 &  496  &   14.1 & 4.67 & 14.8 & 4.9 \\
18511+0146  &	   29.3  &  0.55 & 10.23  &    959 &  137  &  3.89 & 1.29 & 2.3 & 0.8 \\
18517+0437  &	     29  &  0.41 & 7.2$^{{\bf 1}}$  &  336  &  87  & 2.47 & 0.8 & 2.0 & 0.6 \\
19092+0841  &	   11.7  & 0.25 &  5.57  &   426 &  283  &  8.05 & 2.7 & 10.3 & 3.4 \\
20126+4104  &	     18  & 0.15 &  21.9$^{{\bf 2}}$  &   504 &  982  &  28.0 & 9.2 & 61.2 & 20.2 \\
20216+4107  &	   22.7  & 0.19 &  6.1$^{{\bf 2}}$  &    160 &  196   &  5.58 & 1.84 & 9.7 & 3.2 \\
20343+4129  &	     26  &  0.18 &  19$^{{\bf 2}}$  &    426 &   586 &  16.7 & 5.51 & 30.7 & 10.3 \\
22172+5549  &	   15.7  &  0.18 &  3.47   &   239 &   307   &  8.73 & 2.88 & 15.5 & 5.2 \\
\hline
\end{tabular}
\end{center}
$^{\bf 1}$ = from Beuther et al.~(\cite{beuther}), derived from the 1.2~mm continuum \\
$^{\bf 2}$ = from Williams et al.~(\cite{williams}) \\
$^\dagger$ = from $M_{\rm cont}$ \\
$^{\dagger\dagger}$ = from $A_{\rm v}$ \\
\end{table*}

\subsection{\H\ and \D\ column densitites and deuterium fractionation}
\label{sec_deut}

The deuterium fractionation, $D_{\rm frac}$, can be derived by determining the ratio
of the column density of a \mbox{hydrogen-bearing} molecule and its 
deuterated isotopologue. 
It is important to point out that in cold and dense clouds the molecular
species used to compute $D_{\rm frac}$ can be 
affected by depletion. For example, HCO$^{+}$ and H$_2$CO and their 
deuterated counterparts are highly
depleted in the high-density nucleus of low-mass starless cores (e.g. Carey et
al.~\cite{carey}). Therefore, the deuterium fractionation derived from these 
species is representative only of an outer, lower--density shell. 
On the other hand, it has been well established that
\H\ and \D\ do not freeze--out even in the densest 
portions of low-mass molecular clouds (e.g. Caselli et 
al.~\cite{casellia};~\cite{casellib}; 
Crapsi et al.~\cite{crapsi}). Therefore these species should give
an estimate of the deuterium fractionation representative also
of the densest regions of our sources.

We have derived the \H\ and \D\ column densities, $N({\rm N_2H^+})$ and
$N({\rm N_2D^+})$, following the method
outlined in the Appendix of Caselli et al.~(\cite{casellib}), which
assumes a constant excitation temperature, \Tex .
% Following Crapsi et al.~(\cite{crapsi}), 
The values of \Tex\ for the \H\ (1--0) lines have been derived from the 
hyperfine fitting procedure described in Sect.~\ref{obsn2h}. For 
the optically thin lines, for which the fitting 
procedure cannot provide \Tex , we have assumed 
\Tex = $T_{\rm kin}$ (Col.~5 of Table~\ref{tab_ref}) as already
pointed out in Sect.~\ref{sec_spe}. 
%for lines with well-constrained opacity,
%(i.e. with ratio between $\tau$ and its uncertainty $\sigma_{\tau}$ larger
%than 3). For
%the others, we have assumed optically thin lines and used \Tex = 4 K, 
For the \D\ (2--1) line, for which we do not have good
estimates of the optical depth, we have assumed the excitation
temperature of the \H\ (1--0) line.

The results obtained are listed in Table~\ref{tab_dfrac}: the \H\
column densities are of the order of a few $\sim 10^{13}$ \cmq ,
while the \D\ column densities are a few $\sim 10^{11}$ \cmq . The
values of \Dfrac\ are between 0.004 and 0.02, with an average
value of $\sim 0.015$. This value
is $\sim3$ orders of magnitude larger than the cosmic 'average' value of 
$\sim 10^{-5}$ (Oliveira et al., 2003), and close to that
found by Crapsi et al.~(2005) in their sample of low-mass starless cores,
indicating that also the physical conditions of the gas
responsible for the \D\ emission are similar.
%(i.e. $T\sim 10$ K and $n_{\rm H_{2}}\sim 10^{6}$\cmc\ ) are similar.

We have also computed the \H\ abundance relative to H$_2$ by dividing
$N({\rm N_2H^+})$ by the molecular hydrogen column densities listed in 
Col.~8 of Table~\ref{tab_dust} (i.e. the estimates derived from $A_{\rm v}$).
The average value is $\sim 2.05\times 10^{-10}$, which is in excellent
agreement with the $2\times 10^{-10}$ found by Caselli et 
al.~(\cite{casellic}) in a sample of 18 low-mass starless cores. 

Finally, the \H\ and \D\ column densities 
have been computed using the LVG program described in the
Appendix C of Crapsi et al.~(\cite{crapsi}), with which
we have also estimated the $n_{\rm H_2}$ volume densities.
The results are given in Table~\ref{tab_lvg}. 
We note that both the \H\ and \D\ column densities are in 
good agreement with the estimates given in Table~\ref{tab_dfrac}. 
It is interesting to compare also the H$_2$ volume density 
estimates given in Tables~\ref{tab_dust} and
\ref{tab_lvg}. The values obtained in LVG approximation
are on average $\sim 2$ times larger than the estimates 
given in Col.~9 of Table~\ref{tab_dust}, and $\sim 7$ times larger 
than those listed in Col.~10 of Table~\ref{tab_dust}.
However, it must be noted that the sources with the higher
discrepancy are \dcu\ and \dcd , for which the
density estimates have big uncertainties: \dcu\ is the most distant 
of our sources, while for \dcd\ we have derived the flux
at 850 $\mu$m from an extrapolation of the 1.2~mm flux. If we
neglect these two sources, the average ratio between the
density estimates from Col.~10 of Table~\ref{tab_dust}
and those in Table~\ref{tab_lvg} becomes $\sim 4$. Taking into account
that the densities derived in the LVG approximation 
are affected by large uncertainties (even 1 order
of magnitude), we conclude that for the sources with more
accurate measurements the different density
estimates are consistent within the uncertainties, confirming
that the \H\ molecule is a good H$_2$ density tracer.
%and a factor of $\sim 10$ larger than 
%those obtained by Molinari et al.~(\cite{mol00}) and 
%Beuther et al.~(\cite{beuther}). 

Crapsi et al.~(\cite{crapsi}) have
found that the H$_2$ volume densities derived from the LVG code
are systematically lower than those obtained from dust, 
suggesting a possible partial depletion of \H\ in the inner
nucleus of their starless cores. However, this nucleus has
a diameter of $\sim$2500 A.U. (see e.g. Caselli et 
al.~\cite{caselli05}), whose contribution is negligile
at a distance of some kiloparsecs with our angular resolution. 
Also, we have to consider that the H$_2$ volume
densities derived from dust assume an isothermal cloud, which
could not be a good assumption for our sources.

\begin{table*}
\caption{\H\ and \D\ total column densities ($N({\rm N_2H^+})$ and $N({\rm N_2D^+})$),
deuterium fractionation ($D_{\rm frac}$), and \H\ chemical abundance relative to H$_2$
($X_{\rm N_2H^+}$).
$X_{\rm N_2H^+}$ has been calculated from the molecular hydrogen column density
given in Col.~8 of Table~\ref{tab_dust}.}
\label{tab_dfrac}
\begin{center}
\begin{tabular}{ccccc}
\hline \hline
source & $N({\rm N_2H^+})$ & $N({\rm N_2D^+})$ & $D_{\rm frac}$ & $X_{\rm N_2H^+}$ \\
       & ($\times 10^{13}$\cmq )  & ($\times 10^{11}$\cmq )  &  &  ($\times 10^{-10}$) \\
\hline
05345+3157  &   3.14$\pm$0.07 &   3.2$\pm$0.9 &   0.010$\pm$0.007 & 1.5 \\
05373+2349  &   4.34$\pm$0.06 &   2.8$\pm$0.8 &   0.006$\pm$0.001 & 1.9 \\
18144+3157  &   9.75$\pm$0.08 &  $\leq$4.8   &   $\leq$0.004     &  2.1 \\
18511+0146  &   6.47$\pm$0.01 &  2.5$\pm$0.4  &   0.004$\pm$0.001 & 5.0 \\
18517+0437  &   6.48$\pm$0.02 &  $\leq$6.7     &   $\leq$0.01    & 8.1 \\
19092+0841  &   1.9$\pm$0.5 &  $\leq$5.3     &   $\leq$0.028    & 0.7 \\
20126+4104  &   7.62$\pm$0.01 &  4.9$\pm$0.7  &    0.006$\pm$0.001  & 0.8 \\
20216+4107  &   1.4$\pm$0.2 &  2.8$\pm$0.5  &   0.020$\pm$0.007 & 0.8 \\
20343+4129$^{\bf a}$  &   2.2$\pm$0.1 &  4.4$\pm$0.7  &   0.020$\pm$0.005 & 0.4 \\
20343+4129$^{\bf b}$  &   2.9$\pm$0.1 &   5.4$\pm$0.8  &   0.02$\pm$0.01 & 0.5 \\
22172+5549 &   1.226$\pm$0.006  &  2.1$\pm$0.4  &   0.017$\pm$0.004 & 0.4 \\
\hline
\end{tabular}
\end{center}
\footnotesize
$^{\bf a}$ = peak at ($-$8\arcsec ,4\arcsec ) \\
$^{\bf b}$ = peak at (13\arcsec ,0\arcsec ) \\
\normalsize
\end{table*}

\begin{table*}
\caption{H$_2$ volume densities and \H\ and \D\ column densities, derived
in the LVG approximation.}
\label{tab_lvg}
\begin{center}
\begin{tabular}{cccccc}
\hline \hline
    &  \multicolumn{2}{c}{\H\ } & & \multicolumn{2}{c}{\D\ } \\
\cline{2-3} \cline{5-6}
 Source   &  $n_{\rm H_2}$ & $N({\rm N_2H^+})$ & &  $n_{\rm H_2}$ &  $N({\rm N_2D^+})$ \\
          & ($\times 10^{6}$ \cmc ) & ($\times 10^{13}$ \cmq ) & & ($\times 10^{6}$ \cmc ) & ($\times 10^{11}$ \cmq ) \\
\hline
05345+3157 &  2.4  &  3.2 & & 8.0  &   8.6 \\
05373+2349  &  2.8  &  5.0 & & 9.0 &  6.7 \\
18144+3157 &  1.6  &  11 & & -- & --  \\
18511+0146 &  1.0  &  3.9 & & -- & --  \\
18517+0437 &  1.7  &  3.5 & & -- & -- \\
19092+0841 &   1.3  &  7.2 & & -- & -- \\
20126+4104 &   2.6  &  8.6 & &  6.4 &  8.0 \\
20216+4107 &  1.5  &  1.1 & & 1.0 &  6.0 \\
20343+4129$^{\bf a}$ &  0.4 & 4.0 &  &  3.0 &  2.9  \\
20343+4129$^{\bf b}$ &  1.0 &  9.6  & & 6.0 & 30 \\
22172+5549 &   4.0  &    1.4 & & -- & -- \\
\hline
\end{tabular}
\end{center}
$^{\bf a}$ = peak at ($-$8\arcsec ,4\arcsec ) \\
$^{\bf b}$ = peak at (13\arcsec ,0\arcsec ) \\
\end{table*}

\subsection{CO depletion factor}
\label{co_dep}

From the \CIII\ maps we give an estimate of the integrated CO depletion factor, $f_{\rm D}$,
defined as the ratio between the 'expected' abundance of CO relative to H$_2$,
$X_{\rm CO}^{E}$, and the 'observed' value $X_{\rm CO}^{O}$:
\begin{equation}
f_{\rm D}=\frac{X_{\rm CO}^{E}}{X_{\rm CO}^{O}}\;\;,
\end{equation}
where $X_{\rm CO}^{O}$ is the ratio between the observed CO column 
density and the observed H$_2$ column density. 

The \CIII\ column density has been derived from the integrated line 
intensity at the dust peak, assuming optically thin conditions.
This assumption is based on the results of several surveys
of \CIII\ performed in high-mass star formation regions with
comparable mass and density (e.g. Hofner et al.~\cite{hofner};
Fontani et al.~\cite{fonta05}). Under this assumption, and
using the Raleigh-Jeans approximation (valid for our frequencies), 
one can demonstrate that the
column density of the upper level, $J$, is related to the 
integrated line intensity $I({\rm C^{17}O})$ according to:
\begin{equation}
N_J=1.209\, 10^{3}\,\, \frac{2J+1}{J}\frac{k}{\eta_{\nu} \mu^{2}\nu }I({\rm C^{17}O})
\label{eq_nj}
\end{equation}
where $k$ is the Boltzmann constant, $\nu$ the line rest
frequency, $\eta_{\nu}$ the beam filling factor and $\mu$ the molecule's 
dipole moment ($0.11$ Debye for CO). To compute $\eta_{\nu}$, we have 
used the angular diameters listed
in Table~\ref{tab_ang} derived from \CIII\ for all sources
but \zctqc\ and \dcqq , for which we have taken the diameter of 
the sub-mm continuum. The total column density has been obtained from:
\begin{equation}
N_{\rm C^{17}O}=\frac{N_J}{g_J}Q(T_{\rm ex}){\rm exp}\left(\frac{E_J}{kT_{\rm ex}}\right)
\label{eq_ntot}
\end{equation}
where $g_J$ and $E_J$ are statistical weight (=$2J+1$) and
energy of the upper level, and
$Q(T_{\rm ex})$ is the partition function at the temperature 
$T_{\rm ex}$. We have used as excitation
temperature the kinetic temperatures listed in Col.~4 of
Table~\ref{tab_ref}. 
Then, we have computed the observed \CIII\ abundances 
by dividing $N_{\rm C^{17}O}$ by the H$_2$ column densities
listed in Col.~8 of Table~\ref{tab_dust}, i.e. the values obtained
from $A_{\rm v}$. We decided to use these latters rather than
those derived from $M_{\rm cont}$ because they do not assume 
any gas-to-dust ratio.

The \CIII\ 'expected' abundance has been obtained for
each source taking into account the variation of Carbon and
Oxygen abundances with the distance from the Galactic
Center. Assuming the standard value of 9.5$\times 10^{-5}$ 
for the abundance of the main CO isotopologue in the 
neighbourhood of the solar system (Frerking et 
al.~\cite{frerking}), we have computed the expected CO 
abundance at the Galactocentric distance ($D_{\rm GC}$) of
each source according to the relationship:
\begin{equation}
X_{\rm CO}^{E}=9.5\, 10^{-5}{\rm exp}\,(1.105-(0.13\,D_{\rm GC}({\rm kpc})))\;,
\end{equation}
which has been derived according to the abundance gradients
in the Galactic Disk for $^{12}$C/H and $^{16}$O/H listed in Table~1 
of Wilson \& Matteucci~(\cite{wem}), and assuming that the Sun has a
distance of 8.5 kpc from the Galactic Center. Then, following 
Wilson \& Rood~(\cite{wer}), we have assumed that the oxygen
isotope ratio $^{16}$O/$^{18}$O depends on $D_{\rm GC}$ 
according to the relationship $^{16}$O/$^{18}$O=$58.8\times D_{\rm GC}({\rm kpc})+37.1$;
finally, taking the standard ratio $^{18}$O/$^{17}$O=3.52 
(Frerking et al.~\cite{frerking}), we have computed the 
expected abundance of the \CIII , $X_{\rm C^{17}O}^{E}$, for each source 
according to:
\begin{equation}
X_{\rm C^{17}O}^{E}=\frac{X_{\rm CO}^{E}}{3.52\,(58.8\,D_{\rm GC}+37.1)}\;\;\;.
\label{eq_abb}
\end{equation}

The results are listed in Table~\ref{tab_dep}: Cols.~2 and 3 give the
source Galactocentric distance and the corresponding 
$X_{\rm C^{17}O}^{E}$, respectively; Cols.~4, 
5, 6, 7 and 8 list the \CIII\ (2--1) integrated line intensity ($I({\rm C^{17}O})$), 
the beam filling factor ($\eta_{\nu}$), the \CIII\ column 
density ($N({\rm C^{17}O})$), 
the observed \CIII\ abundance ($X_{\rm C^{17}O}^{O}$) and the
CO depletion factor ($f_{\rm D}=X_{\rm C^{17}O}^{E}/X_{\rm C^{17}O}^{O}$). 

Col.~7 of Table~\ref{tab_dep} shows that in 7 out of 9 sources in which we have 
detected \CIII , $f_{\rm D}$ is smaller than 10, while in the
remaining two sources, \vcvs\
and \vtqt , $f_{\rm D}$ is much higher than 10. For \dcu\ we obtain an unusual
value of 0.4. We believe that this is due to the fact that this source 
is the most further away from the Galactic Center ($D_{\rm GC}\sim 12$ kpc), 
and the expected CO abundance at such a distance could be different from that 
calculated from Eq.~(\ref{eq_abb}). 
%Given the large 
%uncertainties on $f_{\rm D}$,
These results indicate that in the majority of our sources
the observed abundance is comparable to or marginally lower
than the expected value. However, the discussion of these results
requires three main comments: first, the values of the
'canonical' CO abundance measured by other authors in
different objects varies by a factor of 2 (see e.g.~Lacy et al.
\cite{lacy}; Alves et al.~\cite{alves}). Second, the integrated
CO depletion factor is an average value along the line of sight.
Therefore, given the large distance to our targets
and the poor angular resolution of our data, 
the effect of not--depleted gas associated with the more
external gaseous envelope of the source can significantly
affect the measured $f_{\rm D}$, which hence is to be taken
as a lower limit.
Finally, the angular resolution of our observations also allow us to
derive only average values of $f_{\rm D}$ over the sources, which 
may have complex structure. 

\begin{table*}
\caption{Parameters used to determine the CO depletion factor: source Galactocentric 
distance ($D_{\rm GC}$), 'expected' \CIII\ abundance ($X_{\rm C^{17}O}^{E}$), integrated 
intensity of \CIII\ at the dust peak position ($I({\rm C^{17}O}))$), beam filling factor
($\eta_{\nu}$), \CIII\ total column density ($N({\rm C^{17}O})$),
observed \CIII\ abundance ($X_{\rm C^{17}O}^{O}$)
and CO depletion factor ($f_{\rm D}$).}
\label{tab_dep}
\begin{center}
\begin{tabular}{cccccccc}
\hline \hline
source & $D_{\rm GC}$ & $X_{\rm C^{17}O}^{E}$ & $I({\rm C^{17}O}))$ & $\eta_{\nu}$ & $N({\rm C^{17}O})$ & $X_{\rm C^{17}O}^{O}$ & $f_{\rm D}$ \\
       & (kpc)  &  ($\times 10^{-8}$) & (K \kms )  &  &  ($\times 10^{15}$\cmq ) & ($\times 10^{-8}$) &  \\
\hline
05345+3157 & 6.38 & 8.62 & 4.10 &  0.34         & 6.04 & 2.9 & 3.0 \\
05373+2349 & 6.96  & 7.38 & $\leq$2.38 & 0.62   & $\leq$1.99 & $\leq$0.9 & $\geq$8.2 \\
18144+3157 & 8.85  & 4.62 & 6.78 & 0.64         & 5.61  &  1.2 &  3.8 \\
18511+0146 & 11.5   & 2.56 & 12.84 & 0.93       & 7.73 & 6.0 & 0.4 \\
18517+0437 & 7.85  &  6.0  & 8.11 & 0.88        & 7.39 & 9.2 & 0.7 \\
19092+0841 & 5.64 &  10.6 & 9.46 & 0.72         & 9.03 &  3.3 & 3.2 \\
20126+4104 &  4.67  & 14.3 & 5.75 & 0.83         & 3.65 &  0.4  & 35.8 \\
20216+4107 &  7.47 & 6.47 & 7.13 & 0.62         & 5.93 & 3.2 & 2.0 \\
20343+4129 &  6.88 & 7.54 & 3.86 & 0.91         & 2.17 &  0.4 & 18.8 \\
22172+5549 &  9.42 & 4.05 & 2.39 & 0.83         & 1.45 &  0.5 & 8.1 \\
\hline
\end{tabular}
\end{center}
\end{table*}

\section{Discussion}
\label{discu}

In the following we discuss the parameters presented in the previous
sections, and compare them with the findings from the literature.

\subsection{Deuterium fractionation and CO depletion}
\label{df_dep}

Theoretical models 
predict that the deuterium fractionation is correlated to
the amount of CO depletion (Caselli et al.~\cite{casellib}; 
Aikawa et al.~\cite{aikawa}). Observations of low-mass 
objects have partially confirmed this theoretical prediction. 
Bacmann et al.~(\cite{bacmann}) have found a good correlation
between $f_{\rm D}$ and $D_{\rm frac}$ (derived from D$_2$CO/H$_2$CO)
in 5 pre--stellar low-mass cores. More recently, Crapsi et 
al.~(\cite{crapsi}) have 
obtained $D_{\rm frac}$ using ${\rm N_2D^+/N_2H^+}$ in a sample of 31 
low-mass starless cores, and they have shown that cores with higher 
$D_{\rm frac}$ have also higher $f_{\rm D}$. 
%However, such a relation seem less evident for values of $f_{\rm D}\leq 10$.

\begin{figure*}
\centerline{\includegraphics[angle=-90,width=10cm]{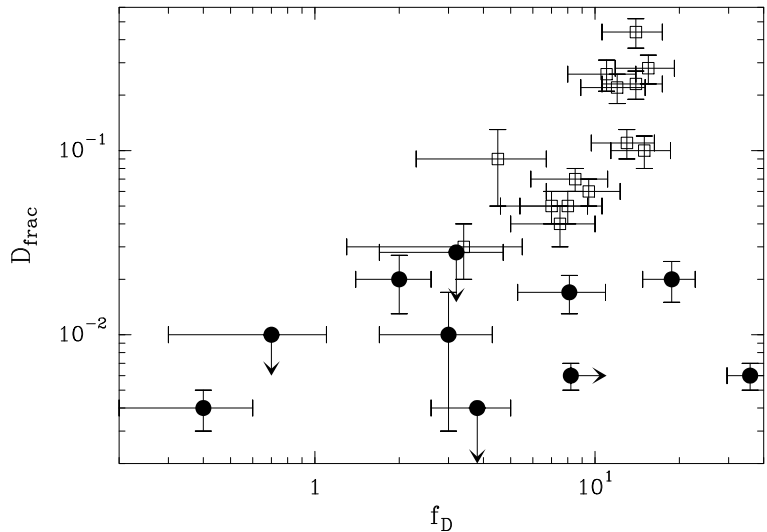}}
\caption{Deuterium fractionation ($D_{\rm frac}$) versus integrated 
CO depletion factor ($f_{\rm D}$) for our sources (filled circles) 
and the low-mass pre--stellar
cores of Crapsi et al.~(\cite{crapsi}) (open squares). 
The arrows indicate the upper limits on
$D_{\rm frac}$ (see Table~\ref{tab_dfrac})
or the lower limit on $f_{\rm D}$ (see Table~\ref{tab_dep}).}
\label{correl1}
\end{figure*}

In order to check if such a correlation exists also in high-mass objects, 
we have added to Fig.~5 of Crapsi et al.~(\cite{crapsi}), in which 
$D_{\rm frac}$ is plotted against the CO
depletion factor, the values obtained from our high-mass objects. The
result is shown in Fig.~\ref{correl1}: we can note that
all the objects of our sample have $D_{\rm frac}$ smaller than 
those of the Crapsi et al.~sample, and nearly half of them
have also smaller $f_{\rm D}$.
%for sources with $f_{\rm D}$ smaller than 10 (among which 8 of our sources,
%considering also the upper limits), there is not a significant 
%difference in the values of $D_{\rm frac}$ between high-mass and
%low-mass objects. On the other hand, the two 
%high-mass objects with $f_{\rm D}> 10$, i.e. \vcvs\
%and \vtqt , have $D_{\rm frac}$ smaller
%than those of the Crapsi et al.~(\cite{crapsi})
%sample with comparable $f_{\rm D}$. However, only \vcvs\ has 
%a value of $f_{\rm D}$ significantly different from the others.
The plot also indicates that the correlation found by
Crapsi et al.~(\cite{crapsi}) and Bacmann et al.~(\cite{bacmann})
for low-mass sources cannot be extended to the sources of our 
sample. In fact, $D_{\rm frac}$ does not show any
dependence from $f_{\rm D}$.

Neverthless, when discussing these results we have to keep in mind
several caveats. First, the sources of Crapsi et al.~(\cite{crapsi}) are 
embedded in low-mass molecular clouds $\sim 100 - 200$ pc away from 
the Sun, therefore their estimates of the deuterium fractionation
and the CO depletion factor,
obtained with the same angular resolution, are less 
affected than ours by the contribution of the non-deuterated
and non-depleted gas along the line of sight associated with the
molecular envelope of the source. 
Additionally, we are comparing a sample of well known starless cores, most
of which are in the {\it pre--stellar} phase,
with a sample of {\it high-mass protostellar} candidates, in which the
heating produced by the forming protostar may affect both deuteration and
CO depletion. Observations of deuterated molecules
in low-mass protostars have shown that the deuterium fractionation is
similar to the values found in pre--stellar cores, as shown
for example by Loinard et al.~(\cite{loinard}), Hatchell~(\cite{hatchell03}),
and Parise et al.~(\cite{parise}), 
who have observed formaldehyde, ammonia and methanol towards
several low--mass protostars. However, these values
are relative to the gaseous envelope in which the low--mass protostars are born, 
which is thought to maintain the initial conditions of the star formation process,
while the heating produced by a forming {\it high-mass} protostar is
expected to dramatically push down the deuteration and the CO depletion of the
molecular environment (see e.g. Turner~\cite{turner}). Finally, as already mentioned in 
Sect.~\ref{co_dep}, the angular resolution of the maps allow us to 
derive only {\it average} values of $D_{\rm frac}$ and $f_{\rm D}$ over 
the sources, which in principle have
complex morphology and may host objects in different 
evolutionary stages (see e.g Kurtz et al.~\cite{kurtz}).
The origin of such cold gas in our sources is thus unclear.
We further discuss this point in Sect.~\ref{where}.

\subsection{Effect of temperature on deuterium fractionation and
CO depletion}
\label{ch_mod}

The deuterium fractionation and the CO depletion factor in a molecular 
core are related to its physical properties. In particular, 
in sources with comparable gas density, $D_{\rm frac}$ and $f_{\rm D}$ 
should be larger in colder clouds. 
%In Fig.~\ref{correl2} we show $D_{\rm frac}$ 
%(panel {\bf a}) and $f_{\rm D}$ (panel {\bf b}) as a function of the gas
%temperature derived from \AMM\ observations (Molinari et 
%al.~\cite{mol96}; Sridharan et al.~\cite{sridharan}). 
%Both panels in Fig.~\ref{correl2} do not indicate a tight correlation
%between $D_{\rm frac}$ and $f_{\rm D}$ and the gas temperature. 
%Neverthless, Panel {\bf b} indicates that in the sources with gas 
%temperature higher than $\sim 25$ K, CO is not depleted,
%as expected from theory, while for those sources with gas 
%temperature around 20 K, the CO depletion can vary from $\sim 3$
%to 22. As already pointed out in Sect.~\ref{df_dep}, when discussing
%these quantities we have to keep in mind that the
%angular resolution of our observations allow us to
%derive only average values over the source, while a
%correlation between $D_{\rm frac}$, $f_{\rm D}$ and gas temperature
%is expected only in the starless cores eventually embedded
%inside the large-scale cloud. 
In this section, we will describe a simple chemical model which has been 
used to interpret the present
observational results on the depletion factor, the deuterium fractionation
and the gas temperature. 
Unlike low mass cores, the structure of massive cores cannot be derived in 
detail with current observations.  The physical properties listed in Table
~\ref{tab_ref} and \ref{tab_dust} are average values within telescope beams 
whose angular sizes are 
comparable to source sizes (compare HPBW in Table 1 with $\theta_{\rm cont}$
in Table~\ref{tab_dust}).  Therefore, any attempt of modeling massive cores as smooth 
objects with densities and temperature gradients similar to those found in 
low-mass cores is subject to large uncertainties, also considering that massive
star forming regions are probably clumpy, so that dense and cold material may 
be confined in small regions with filling factors significantly smaller than 
unity.  

For these reasons, we decided to use a chemical model simpler than the one 
used for low mass cores (see e.g. Crapsi et al. 2005). In particular, the 
detailed physical structure is neglected and the clouds are treated as 
homogeneous objects with density and temperature equal to the average 
values from column 10 of Table~\ref{tab_dust} and column 5 of 
Table~\ref{tab_ref}, respectively. As already pointed out in Sect.~\ref{sec_dust},
we used the gas 
temperature instead of the dust temperature to compare with our model 
predictions, given that the latter has been determined using also the IRAS measurements, 
which are not sensitive to temperatures lower than 30 K (unlike the 
data used to determine $T_{\rm gas}$), and it is thus expected
to overestimate the dust temperature throughout the cloud. If clumps denser and 
colder than the average values are indeed present, our calculation is expected
to underestimate the observed CO depletion factor and the deuterium 
fractionation in our sources, unless the filling factor is significantly less than 
unity.  

Other simplifications include the determination of the electron fraction, 
which is now simply assumed to be given by (following McKee 1989):

\begin{eqnarray}
x(e)=1.3 \times 10^{-5} \times n({\rm H_2})^{-0.5}
\end{eqnarray}

Dust grains (all assumed to be negatively charged
\footnote{Although this is probably true within a factor of 2 (e.g. 
Flower \& Pineau des For\^ets~2003), this 
assumption does not affect our conclusions.}) are distributed in size 
according to Mathis, Rampl \& Nordsieck (1977; hereafter MRN) (but with the 
minimum size value enlarged by a factor of 10, as in the case of low--mass 
cores; if we assume the MRN value, conclusions do not appreciately change
in the range of physical parameters considered here). We also included in the 
model all the multiply deuterated forms of H$_3^+$ and their destruction 
via dissociative recombination with electrons, recombination onto negatively
charged grains, and through reactions with CO, O and N$_2$ has been taken 
into account. The freeze--out 
of CO, O and N$_2$ is balanced by thermal (see e.g. Hasegawa et al. 1992)
and non--thermal (cosmic--ray impulsive heating; Hasegawa \& Herbst 1993) 
desorption.  In analogy with similar work done on low--mass cores, the 
binding energies used are those for CO and N$_2$ onto H$_2$O (see \"Oberg 
et al. 2005), and the oxygen binding energy is fixed at 
800 K (Tielens \& Allamandola~\cite{tea}). The cosmic--ray
ionization rate used here is 3$\times$10$^{-17}$ s$^{-1}$, an appropriate
value for high--mass star forming regions (e.g. van der Tak \& van Dishoeck
2000).  

The model has been run for different temperature values ($T_{\rm gas}$ = 
$T_{\rm dust}$ from 10 to 100 K) and volume density fixed at the average 
value of the objects in our sample ($n({\rm H_2}) = 6\times 10^5 {\rm cm^{-3}}$)
.  All models start with undepleted abundances of the neutral 
species and run for a time given by the free--fall time appropriate for the 
chosen density ($t_{\rm ff}$ = 5$\times$10$^4$ yr), which implies that 
chemical and dynamical timescales are assumed to be comparable. 

\begin{figure*}
\centerline{\includegraphics[angle=0,width=10cm]{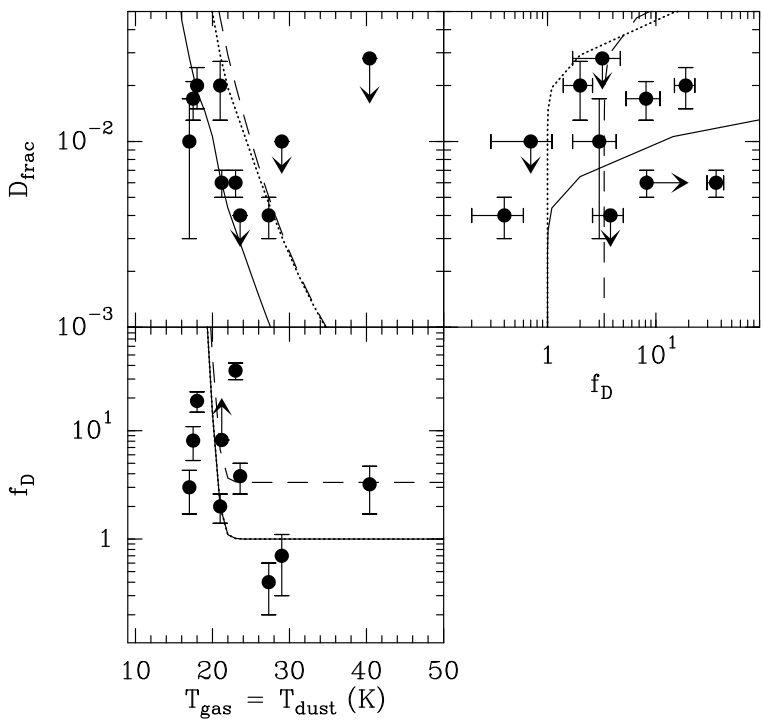}}
\caption{{\bf Top left.}
Deuterium fractionation versus kinetic temperature derived
from NH3 observations. The arrows indicate the deuterium fractionation
upper limits.  The curves represent the prediction of the theoretical
model described in Sect.~4.2.  Solid curve is model prediction when
using the rate coefficients measured by Gerlich et al. (2002) for the
proton-deuteron exchange reactions.  Dotted curve is from the standard
model, with (factor of about 3) larger rate coefficients. Dashed curve
is from the standard model, when 70\% of CO molecules are trapped in
H$_2$O ice.  
\newline
{\bf Bottom.} Same as Top for the integrated CO depletion factor.
\newline
{\bf Top right.} 
Deuterium fractionation versus depletion factor for the same
models and data in the previous two panels.  Note that the presence of trapped CO
does not significantly affect the deuterium fractionation.}
\label{correl2}
\end{figure*}

Figure~\ref{correl2} shows the model predictions for $D_{\rm frac}$, 
i.e. the deuterium fractionation in species such as \H , and the CO 
depletion factor $f_{\rm D}$ versus the gas
(dust) temperature. The thin curves show models where the Gerlich et al.~
(\cite{gerlich}) rate coefficients for the proton--deuteron exchange reactions are used,
whereas the dotted curve is for models with the (factor of ~3) larger
rate coefficients typically used in chemical models (e.g. Roberts et al.~2003, 2004; 
but see Walmsley et al.~2004; Flower et al.~\cite{flower05}). In fact, the
recently measured rate coefficients still suffer some uncertainties, especially 
due to
the importance of the so--called "back reactions" of the deuterated forms of
H$_3^+$ with ortho-H$_2$, which lower the deuteration depending on the unknown
ortho/para H$_2$ ratio.

The data points in Figure~\ref{correl2} suggest that the model is in quite good
agreement with observations for the majority of the objects, despite of the simplicity
of it. Considering
the uncertainties in the rate coefficients and in the gas/dust temperature,
the observed deuterium fractionation is well reproduced in the observed
range of temperatures. In the case of the CO depletion factor, there are
three objects that show too large $f_{\rm D}$ for their adopted temperatures:
IRAS~20126+4107, IRAS~18144-1723, and IRAS~19092+0841. In the case of 
IRAS~20126+4107
and IRAS~19092+0841 we note that the excitation temperature derived from
\H\ (1-0) is indeed close to 10 K, suggesting that in these cases the
deuterated gas may be confined in smaller and colder clumps, compared to
the high mass star forming region where they are embedded. Therefore, lower
dust/gas temperatures (close to 10 K) should be more appropriate.

However, for IRAS~18144-1723, the \H\ (1-0) excitation
temperature is quite large (~26 K), so that some other mechanisms may be
present to maintain a large fraction of CO frozen onto dust grain (unless
the dust temperature is at least 5 K lower than the gas temperature).  One
possibility could be the trapping of CO molecules in H$_2$O ice. Indeed, if
a large fraction of CO is trapped in water ice, dust temperatures
between 30 and 70 K are needed in order to release all the CO back in the
gas phase (Collings et al.~\cite{collings}; Viti et al.~\cite{viti}). To test this
possibility, we run models where a certain fraction of CO molecules were 
assumed to stick
onto dust grains with binding energies appropriate for H2O ice (4820 K;
Sandford \& Allamandola 1990). In the case of IRAS~18144-1723, to reproduce
the observed $f_{\rm D}$ value (3.8) at the measured temperature (23.6 K), 70\% of solid
CO needs to be trapped into water and this is what is shown in Figure~\ref{correl2} 
by the dashed curves, which is also able to reproduce the $f_{\rm D}$ value observed 
in the warmest source (\dznd ). 
However, the CO trapping does not affect the deuterium
fractionation, which is limited by the gas temperature.  We then conclude
that
a significant fraction of CO may be indeed trapped in H$_2$O ice, although
more accurate determination of the dust and gas temperature are needed
to give more quantitative estimates.

%\begin{figure*}
%\centerline{\includegraphics[angle=-90,width=10cm]{correl2_cas.ps}}
%\caption{Same as Fig.~\ref{correl1} but without the
%sources of Crapsi et al.~(\cite{crapsi}). The arrows indicate
%the deuterium fractionation upper limits. The solid line 
%represents the prediction of the theoretical model described in
%Sect.~\ref{ch_mod}.}
%\label{correl3}
%\end{figure*}

\subsection{Where are the pre--stellar cores?}
\label{where}

The most important result of this work is the detection of
\D\ emission in 7 of our sources, with values of deuterium
fractionation close to those found in low-mass starless
cores. In this section we discuss this result,
focusing the attention on the origin and the nature of the cold and
dense gas that gives rise to this emission.

While Crapsi et al.~(\cite{crapsi}) have established that
in their objects the \D\ emission arises from the densest core 
nucleus, the angular resolution of our data does not allow one to
determine the accurate location of this emission. In fact,
as already noted in Sects.~\ref{df_dep} and \ref{ch_mod}, 
it is well known that
high-mass stars typically form in clusters, so that the surroundings
of a massive protostar often show complex morphology (see Molinari et
al.~\cite{mol02}, Cesaroni et al.~\cite{cesa03}, Fontani et 
al.~\cite{fonta1}) and may be fragmented in objects with different 
masses and in different evolutionary stages (see e.g. Kurtz et 
al.~\cite{kurtz}; Fontani et al.~\cite{fonta2}), whose angular
separation can be comparable to or even smaller than the angular
resolution of the observations. 

For this reason, in principle the observed cold and dense gas responsible for the
\D\ emission may be associated with the high-mass protostellar object 
or with another object located very close to it.
In the first case, such gas is located in the most external shell of 
the parental cloud not yet heated up by the high-mass
(proto-)star: a 'record' of the early cold phase.
As a matter of fact, relatively high values of deuterium fractionation
have been found also in some hot cores (e.g. Oloffson~\cite{oloffson},
Turner~\cite{turner}), i.e. the molecular environment
in which massive stars have been recently formed. Such a scenario 
would indicate that the
molecular cloud in which low-mass and high-mass stars are born 
have similar chemical and physical conditions.
In the second case, the \D\ emission is due to one or more molecular 
cores in the pre--stellar phase
located close to the central protostar.
In fact, high--angular resolution observations of one of our targets,
\zctqc , have clearly shown that the molecular gas is fragmented
into several cores (see Molinari et al.~\cite{mol02}), separated on average by 
$\sim 5$\asec -- 10\asec . Such a separation
is smaller than the angular resolution of our data. 
Therefore, the observed \D\ emission might arise from some of
these cores. Given the comparable distances,
in principle this could be the case also for the other sources.
Of course, to solve this problem observations with higher angular 
resolutions are absolutely required.

\section{Conclusions}
\label{conc}

We have observed several rotational lines of \H , \D , \CIII, \DCO\ and
sub-mm continuum emission with the IRAM-30m telescope and the JCMT in 10
high-mass protostellar candidates. Our main goal is to measure
the deuterium fractionation through the ratio 
$N({\rm N_2D^+})/N({\rm N_2H^+})$, and the CO depletion factor
(ratio between 'expected' and observed CO abundance), in
order to shed light on the chemical and physical properties of
the molecular clouds in which high-mass stars are born.
The main results of this study are the following:
\begin{itemize}
\item We have detected \D\ emission in 7 out of 10 sources of our
sample, with values of $N({\rm N_2D^+})/N({\rm N_2H^+})$ between
$\sim 0.004$ and $\sim 0.02$, with an average value of 0.015. This value
is 3 orders of magnitude
larger than the interstellar value for the relative D/H abundance
($\sim 10^{-5}$), and close
to that found by Crapsi et al.~(\cite{crapsi}) in their sample
of low-mass starless cores.
\item For almost all the observed sources the sub-mm continuum map and the
maps of the \H\ (1--0) and (3--2) integrated line intensity show a
similar distribution of the emission, reinforcing the idea that 
\H\ and sub-mm continuum trace the same material. 
%On the other hand, maps of
%the \CIII\ (2--1) line integrated intensity show a different 
%distribution.
\item The CO 'expected' abundances are comparable to or marginally higher than 
the values measured 
from the observations in 8 sources, and much higher in 2 sources
(\vcvs\ and \vtqt ). The median value is 3.2, which is smaller
than that found in low-mass pre--stellar cores. However, the non-depleted 
gas along the line-of-sight may strongly affect these results,
since the source distances are larger than 1 kpc. 
We thus believe that the 'real' CO depletion factors
estimated by us are lower limits. 
\item 
A simple chemical model suggests that the majority of the observed sources
have conditions similar to low-mass pre--stellar cores (large CO depletion
factors and deuterium fractionation). The observed deuterium fractionation
and CO depletion can be reproduced if the
gas and dust temperature are $\sim 20$ K, i.e. close to that observed
by \AMM\ or \ace\ observations and to the \H\ (1--0) excitation temperature.
Although some of the studied sources (IRAS~20126+4104 and IRAS~19092+0841)
may indeed host dense and cold clumps with physical and chemical
characteristics identical to low-mass pre--stellar cores, on average
the present observations are consistent with envelope material recently
heated by the central massive star, suggesting that the studied sources may
have experienced a pre--stellar phase with gas and dust temperatures  quite
similar to those observed in the nearby low-mass cores ($\sim 10$ K), despite the
larger volume densities. It appears that large fractions of CO may be
trapped in H$_2$O ice, although high angular resolution observations are
needed to confirm this.
%However, this last statement can only be confirmed 
%with higher angular resolution observations.   

\end{itemize}

Our results allow us to conclude that most of our sources host
cold and dense gas with
properties similar to those found in low--mass starless cores on the verge
of the gravitational collapse.
The problem to solve now is the location and the origin of this gas.
We propose two scenarios: in the first one, the cold gas is distributed in an
external shell not yet heated up by the high-mass protostellar object,
a remnant of the parental massive starless core. In the second one,
the cold gas is located in cold and dense cores close to the high-mass
protostar but not associated with it.
This scenario is supported by high--angular resolution observations
of some of our objects, which have revealed their clumpy structure.
Clearly, observations at higher angular resolutions will be
very helpful in deciding which solution is correct.

\newpage
\newpage
\newpage
\newpage
\begin{acknowledgements}
We thank C. M. Walmsley for his valuable comments and suggestions.
Antonio Crapsi was supported by a fellowship from the European Research Training
Network "The Origin of Planetary Systems'' (PLANETS, contract number
HPRN-CT-2002-00308) at Leiden Observatory.
\end{acknowledgements} 

\renewcommand{\thefigure}{A-\arabic{figure}}
% redefine the command that creates the equation no.
\setcounter{figure}{0}  % reset counter 
\section*{Appendix: spectra and maps} 

\begin{figure*}
\centerline{\includegraphics[angle=0,width=15cm]{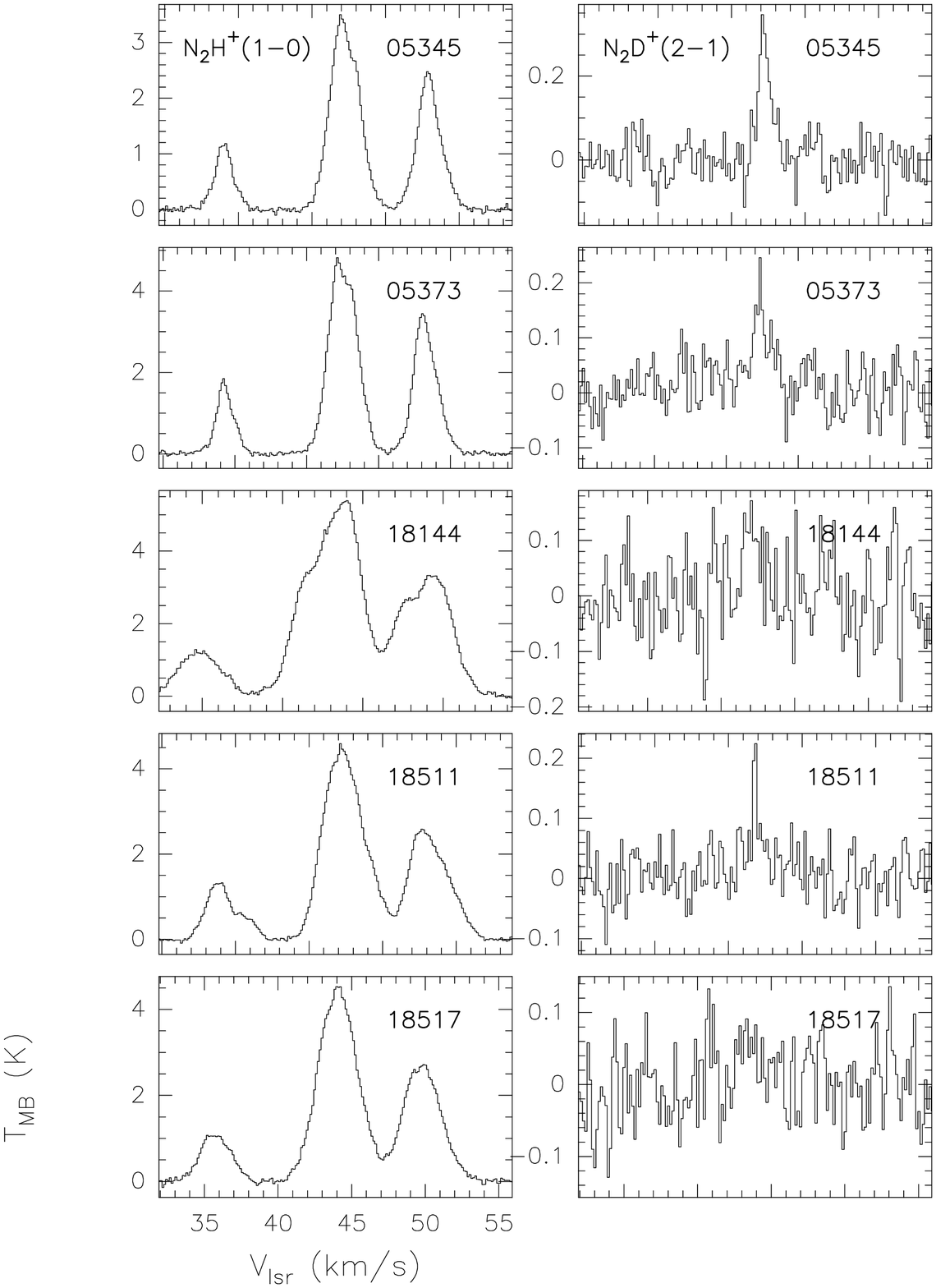}}
\caption{Spectra of the \H\ (1--0) line (left panels) and the \D\ (2--1)
line (right panels) for \zctqc , \zctst , \dcqq , \dcu\ and \dcd\
(from top to bottom).
All spectra have been taken at the position of the \H\ (1--0) line
emission peak.}
\label{fig_spe1}
\end{figure*}

\begin{figure*}
\centerline{\includegraphics[angle=0,width=15cm]{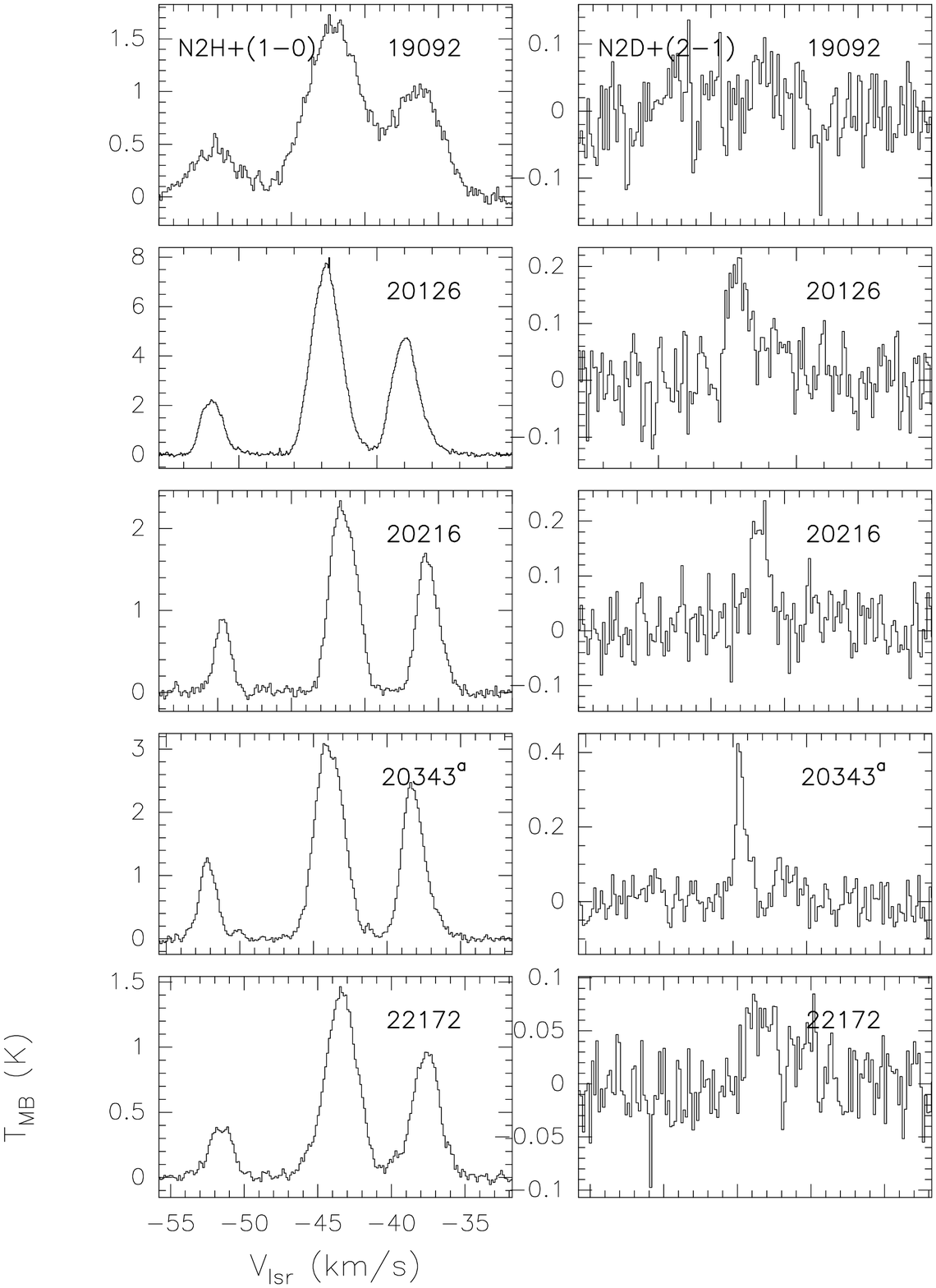}}
\caption{Same as Fig.~\ref{fig_spe1} for \dznd , \vcvs , \vds ,
\vtqt\ and \vdcsd . For \vtqt , we show the spectra obtained towards
position $^{\bf a}$ (see Table~\ref{tab_lin_n2h}).}
\label{fig_spe2}
\end{figure*}

\begin{figure*}
\centerline{\includegraphics[angle=-90,width=17cm]{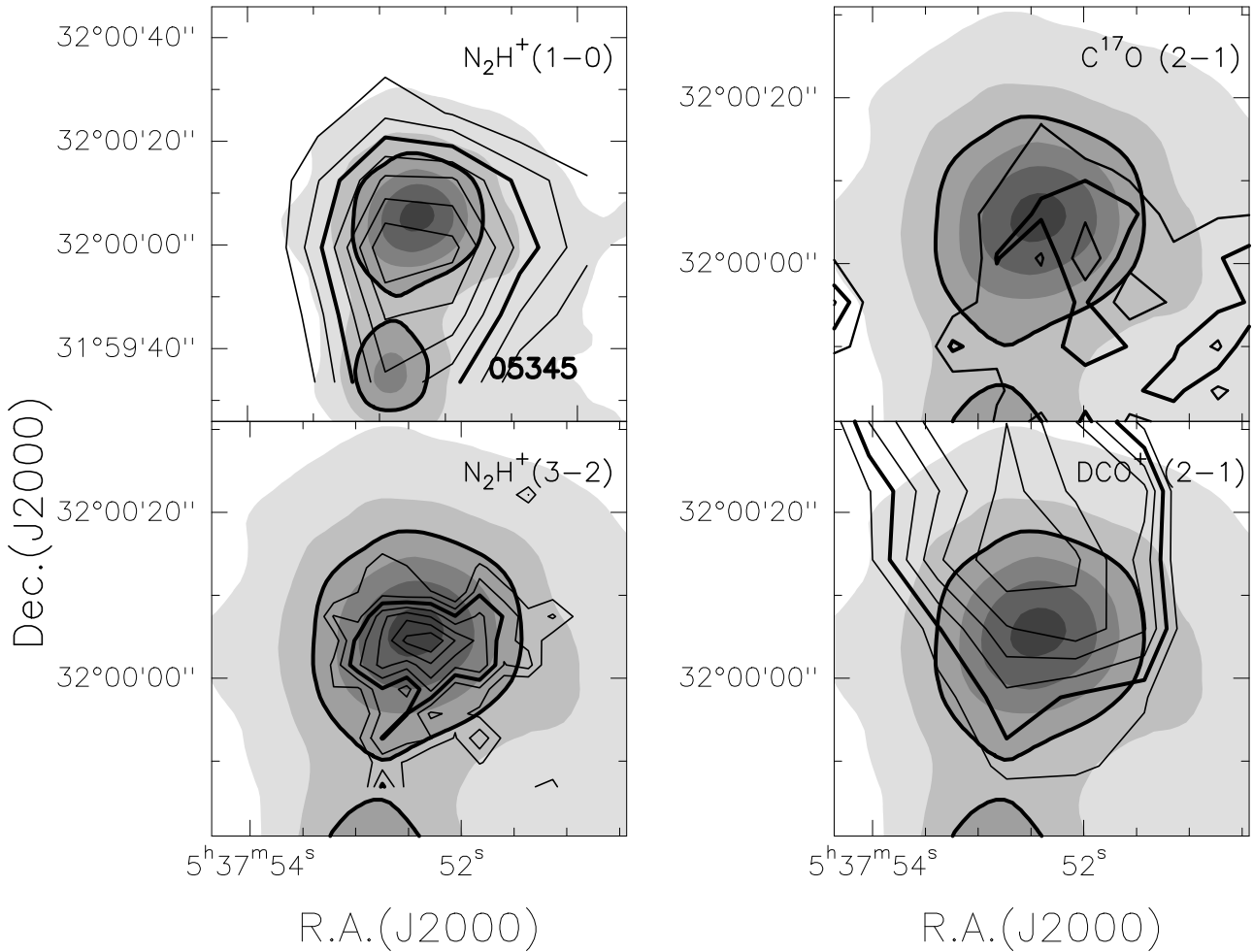}}
\caption{Left panel: plot of the \H\ (1--0) (top panel) and (3--2) (bottom
panel) integrated emission, superimposed on the 850 $\mu$m map (grey-scale), 
obtained with SCUBA towards \zctqc . Right panel: same as left panel for the \CIII\ (2--1)
line (top panel) and \DCO\ (2--1) line (bottom panel).
The levels are:
from 20$\%$ to 95$\%$ of the peak (step 15$\%$) for the SCUBA map; 
from 30$\%$ to 90 $\%$ of the peak (step 10$\%$) for the
\H\ lines; from 40$\%$ to 90$\%$ (step 10$\%$) for the other lines.
The contours at half of the maximum are indicated by thick lines
for each tracer.}
\label{05345_map}
\end{figure*}
\begin{figure*}
\centerline{\includegraphics[angle=-90,width=17cm]{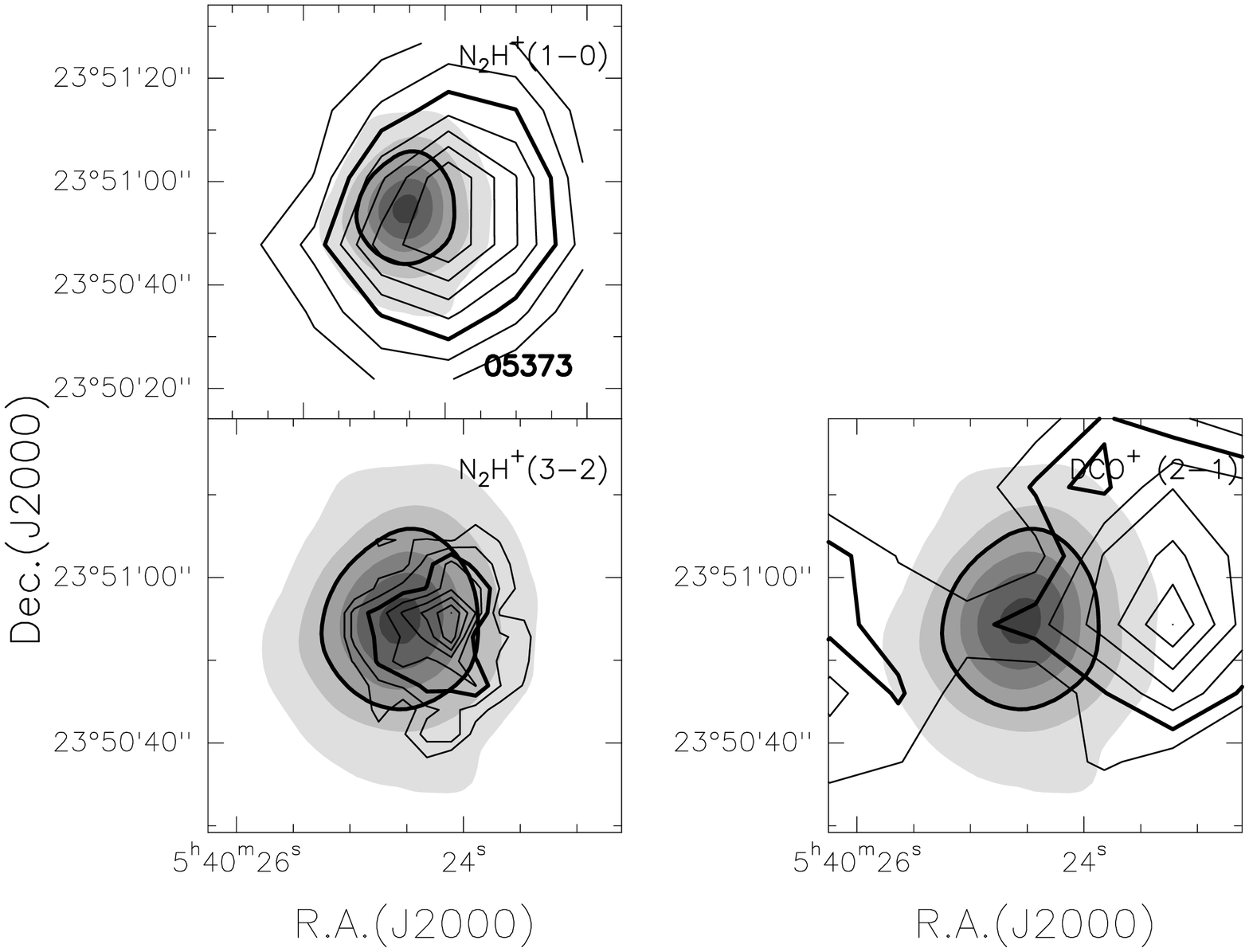}}
\caption{Same as Fig.~\ref{05345_map} for \zctst , which is undetected
in \CIII .}
\label{05373_map}
\end{figure*}
\begin{figure*}
\centerline{\includegraphics[angle=0,width=12cm]{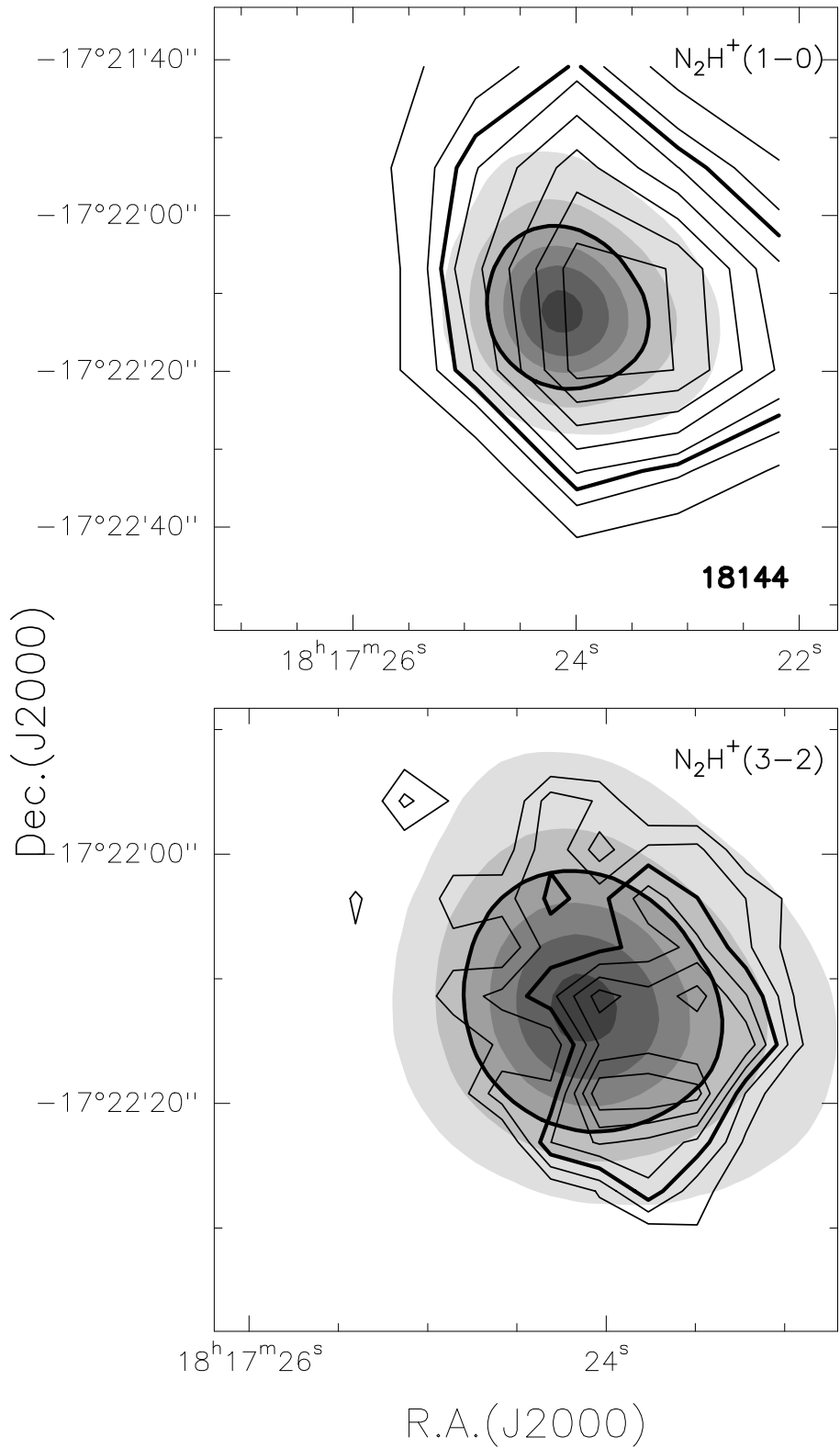}}
\caption{Same as Fig.~\ref{05345_map} for \dcqq . We do not show the \CIII\ 
map because it is too noisy, nor the \DCO\ map because the source
is not detected in this tracer.}
\label{18144_map}
\end{figure*}
\begin{figure*}
\centerline{\includegraphics[angle=-90,width=17cm]{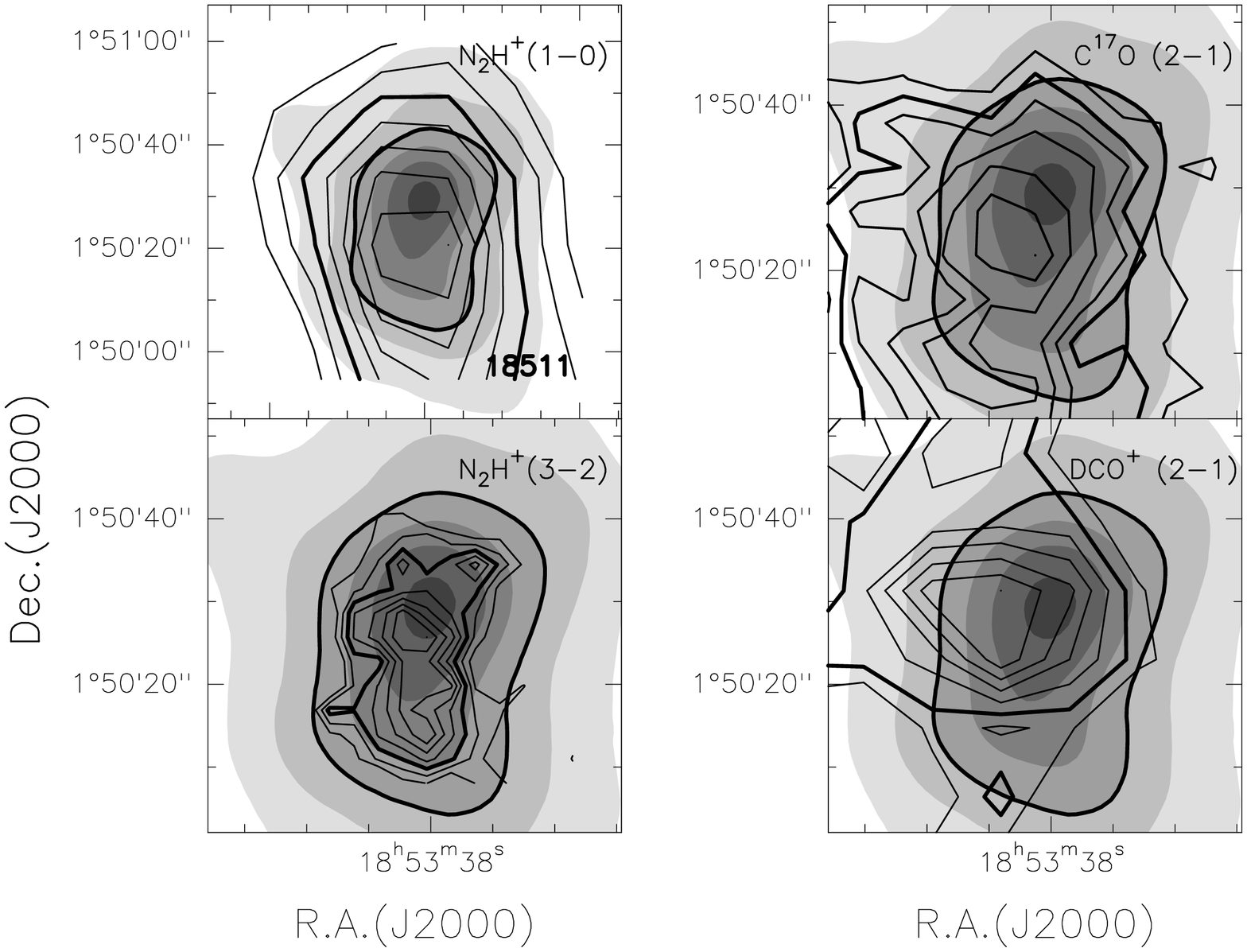}}
\caption{Same as Fig.~\ref{05345_map} for \dcu .}
\label{18511_map}
\end{figure*}
\begin{figure*}
\centerline{\includegraphics[angle=-90,width=17cm]{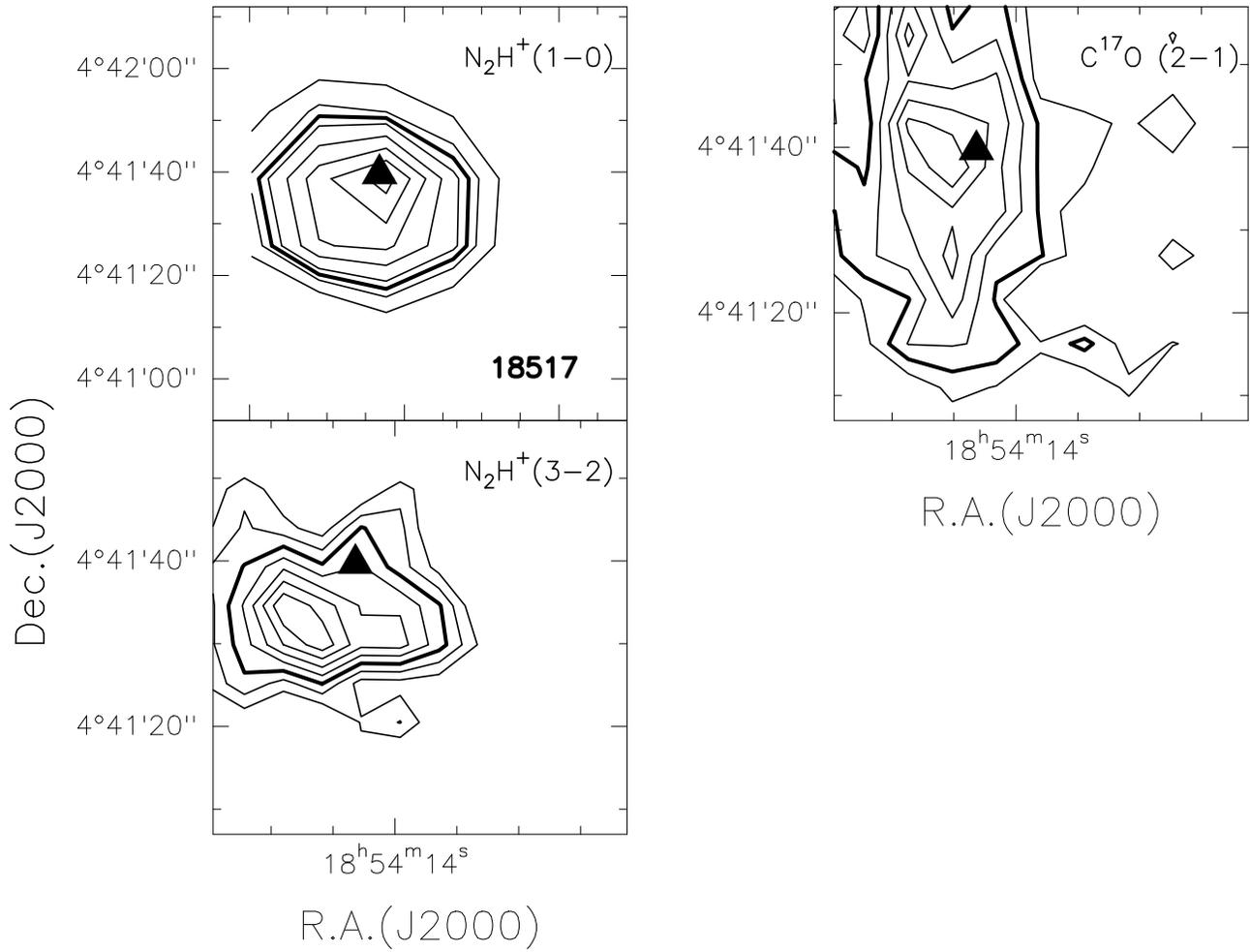}}
\caption{Left panel: integrated \H\ (1--0) (top panel) and
(3--2) (bottom panel) line emission in \dcd . Right panel: integrated
emission of the \CIII\ (2--1) line. In each panel, the filled triangles
at map center indicate the position of the 1.2~mm emission peak
(Beuther et al.~\cite{beuther}).}
\label{18517_map}
\end{figure*}
\begin{figure*}
\centerline{\includegraphics[angle=-90,width=17cm]{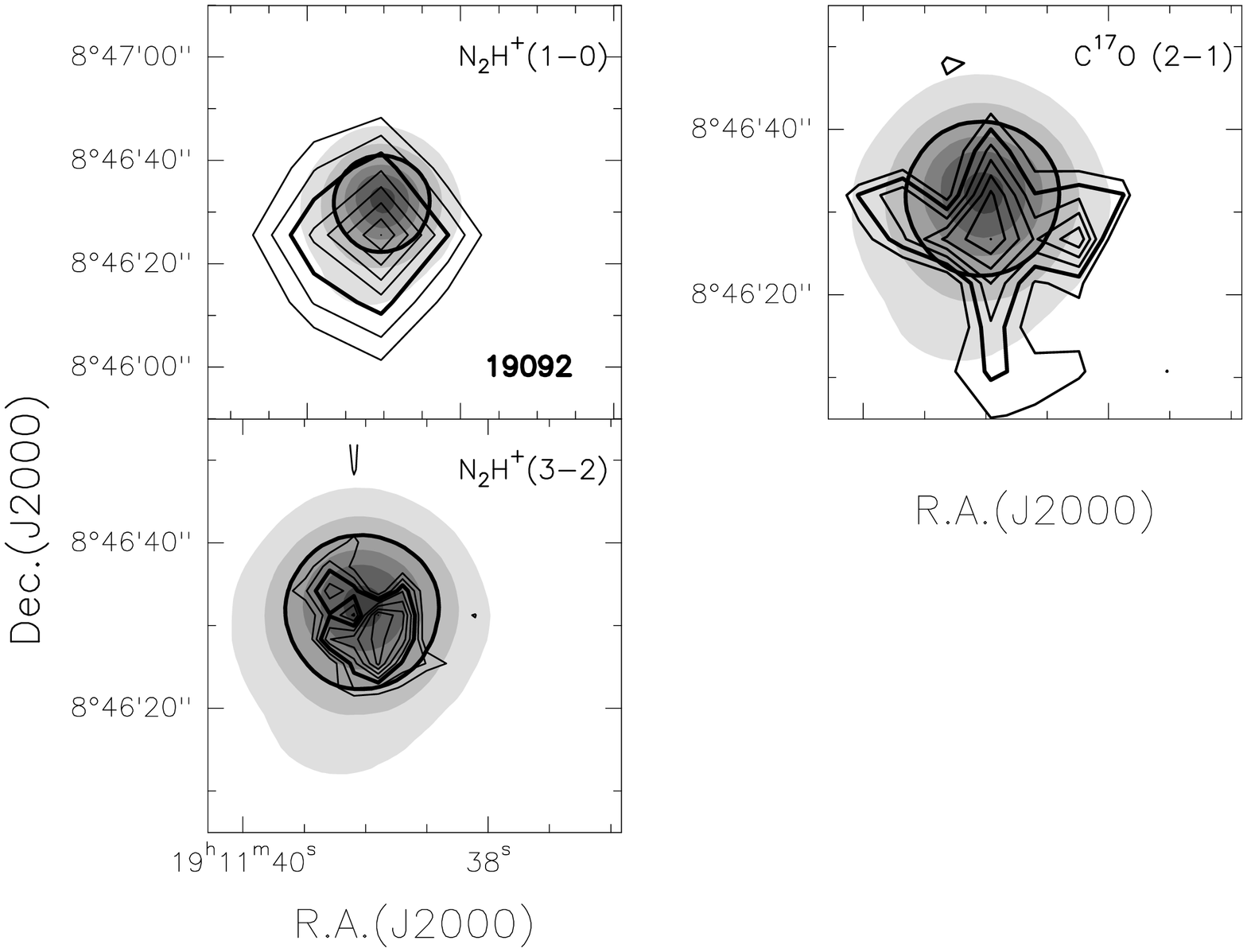}}
\caption{Same as Fig.~\ref{05345_map} for \dznd .}
\label{19092_map}
\end{figure*}
\begin{figure*}
\centerline{\includegraphics[angle=-90,width=17cm]{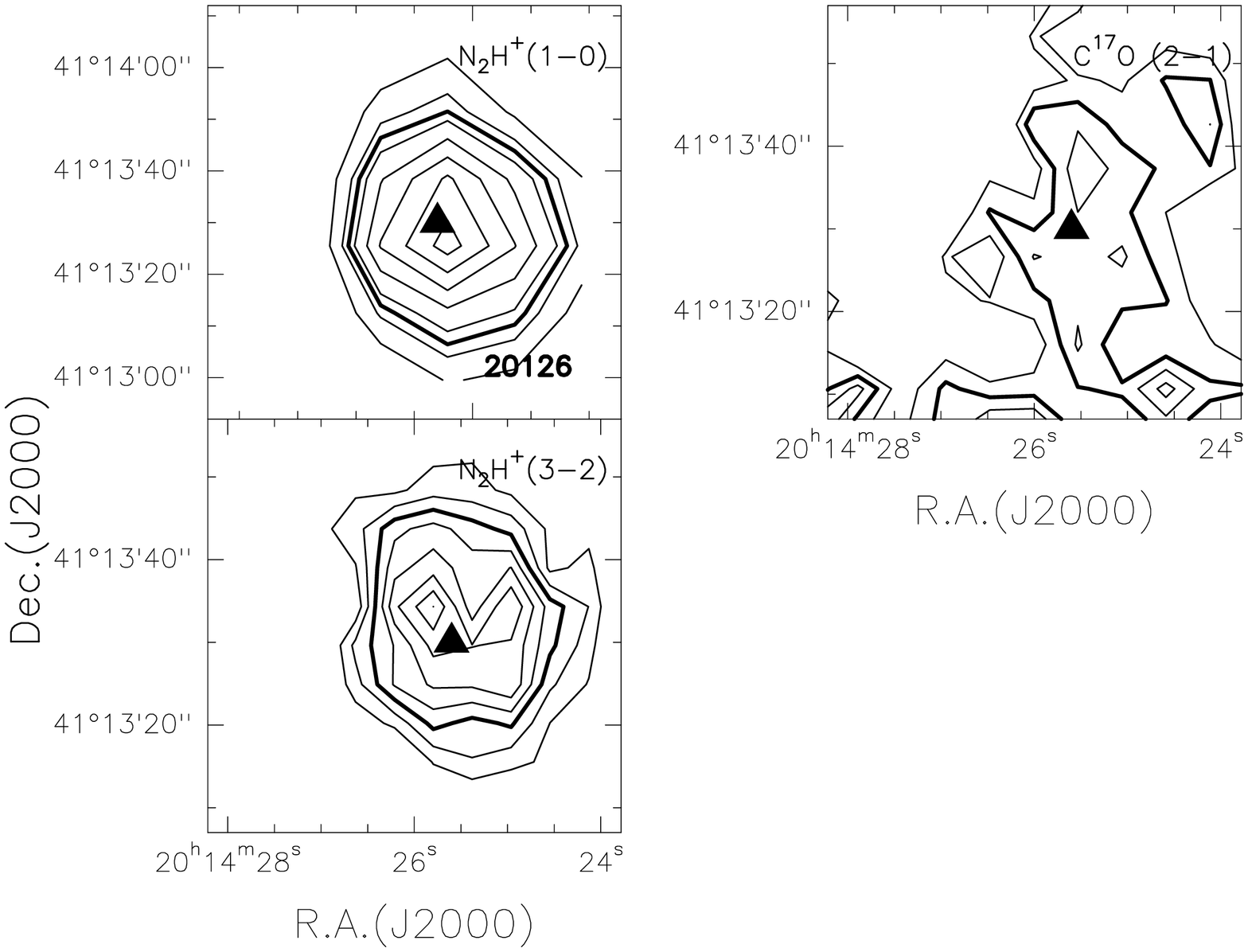}}
\caption{Same as Fig.~\ref{18517_map} for \vcvs . The triangle corresponds
to the position of the 850~$\mu$m emission peak (Williams et al.~\cite{williams}).}
\label{20126_map}
\end{figure*}
\begin{figure*}
\centerline{\includegraphics[angle=-90,width=17cm]{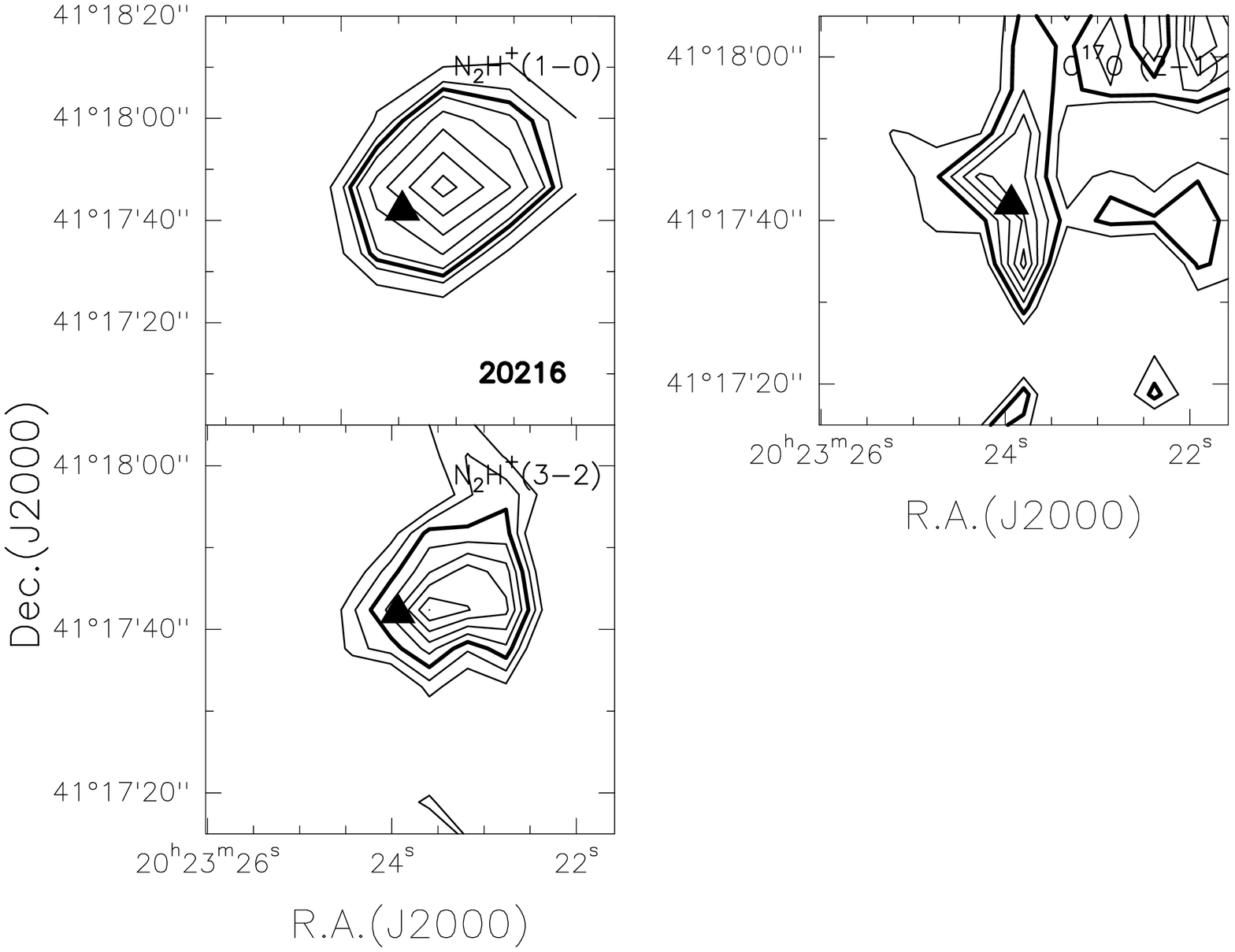}}
\caption{Same as Fig.~\ref{20126_map} for \vds .}
\label{20216_map}
\end{figure*}
\begin{figure*}
\centerline{\includegraphics[angle=-90,width=17cm]{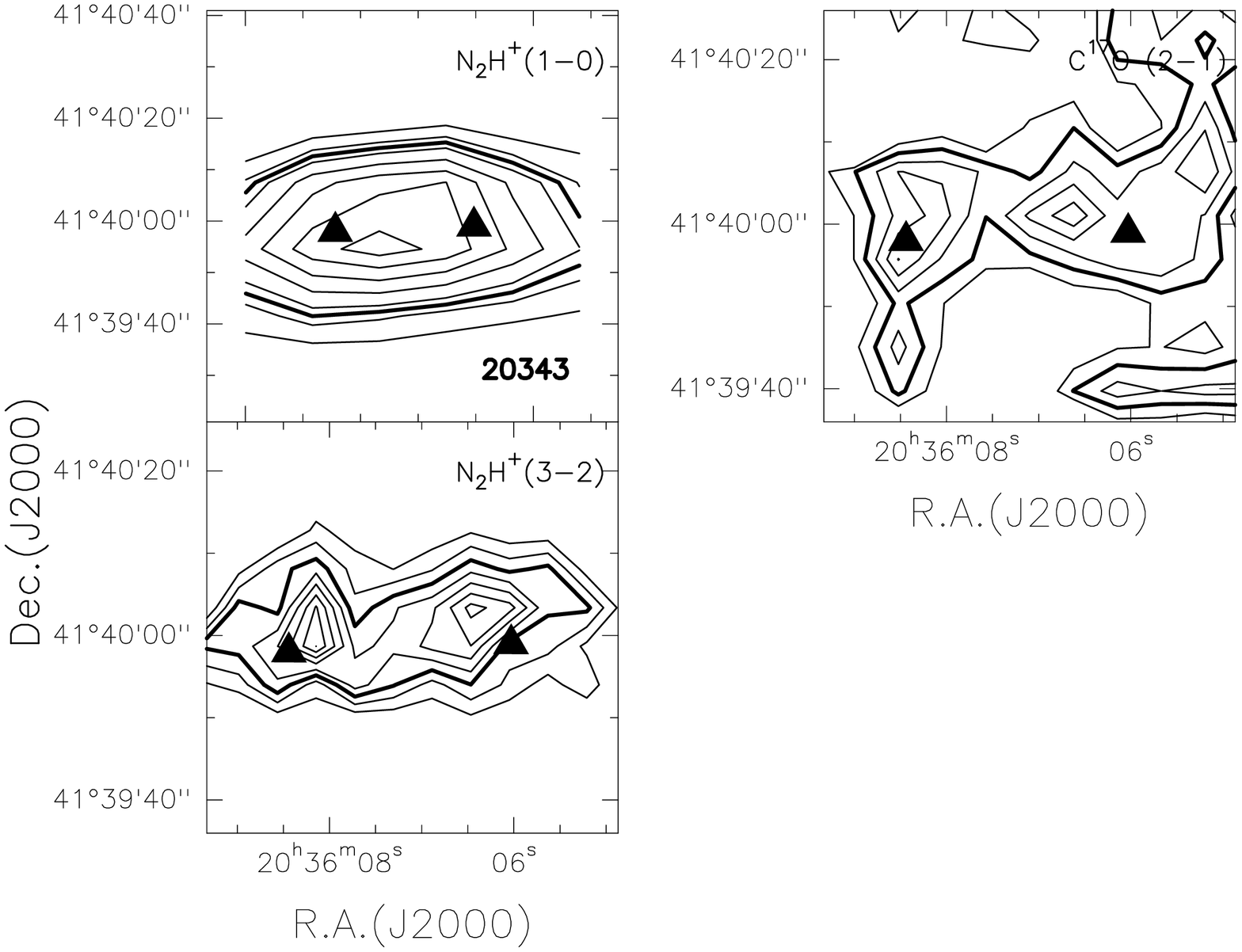}}
\caption{Same as Fig.~\ref{20126_map} for \vtqt . }
\label{20343_map}
\end{figure*}
\begin{figure*}
\centerline{\includegraphics[angle=-90,width=17cm]{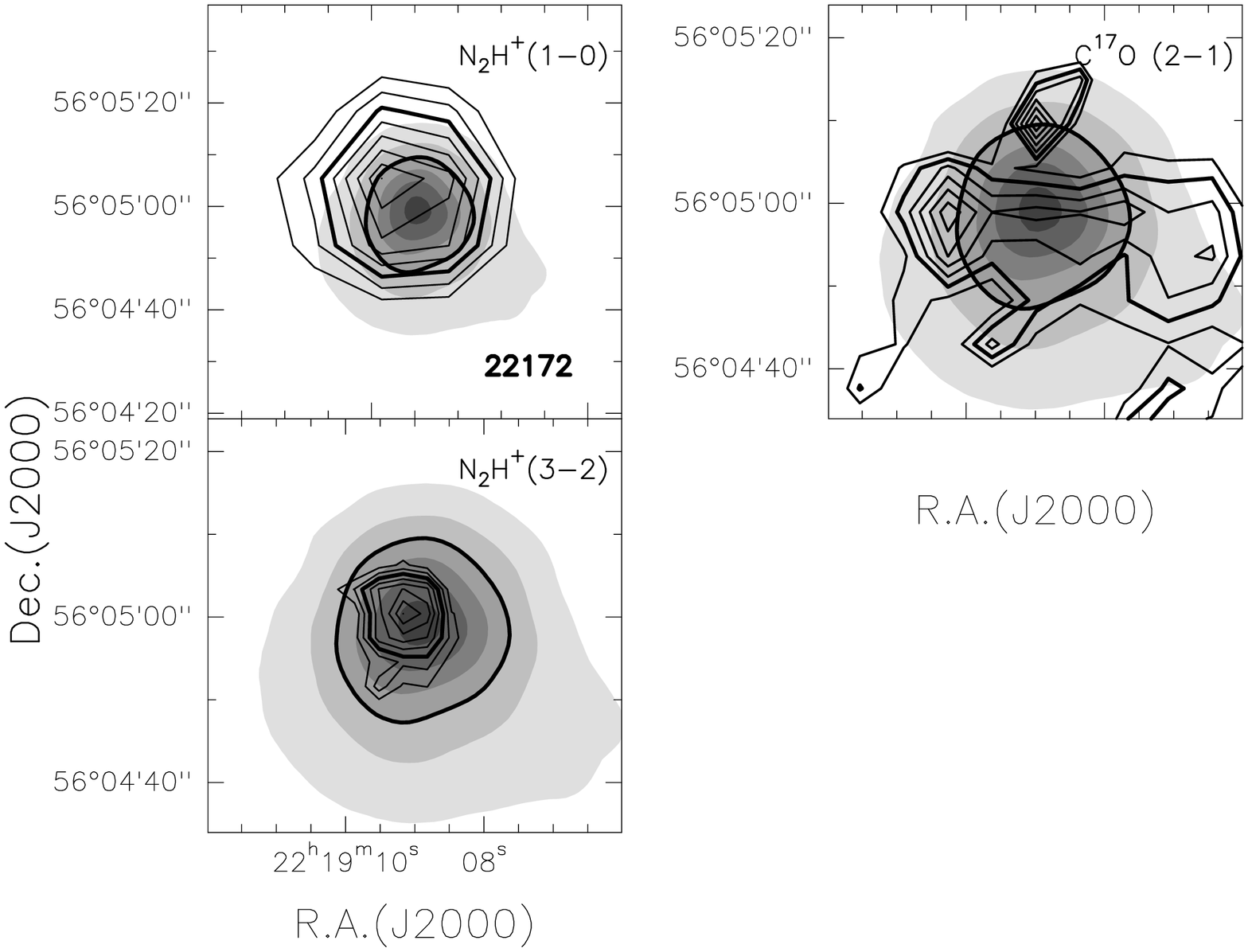}}
\caption{Same as Fig.~\ref{05345_map} for \vdcsd . }
\label{22172_map}
\end{figure*}

{}


\begin{thebibliography}{}
\bibitem[2005]{aikawa}
Aikawa, Y., Herbst, E., Roberts, H., Caselli, P. 2005, ApJ, 620, 330
\bibitem[1999]{alves}
Alves, J., Lada, C.J., Lada, E.A. 1999, ApJ, 515, 265
\bibitem[2003]{bacmann}
Bacmann, A., Lefloch, B., Ceccarelli, C. et al. 2003, ApJ, 585, L55
\bibitem[2006]{beltran}
Beltr\'an, M.T., Brand, J., Cesaroni, R. et al. 2006, A\&A, 2006, 447, 221
\bibitem[2001]{bergin01}
Bergin, E.A., Ciardi, D.R., Lada, C.J., Alves, J., Lada, E.A., 2001, ApJ, 557, 209 
\bibitem[2002]{bergin02} 
Bergin, E.~A., Alves, J., Huard, T., \& Lada, C.~J.\ 2002, ApJ, 570, L101
\bibitem[2002]{beuther}
Beuther, H., Schilke, P., Menten, K.M. et al. 2002, ApJ, 566, 945
\bibitem[2006]{bisschop}
Bisschop, S.E., Fraser, H.J., \"Oberg, K.I., van Dishoek, E.F., Schlemmer, S. 2006, A\&A, 449, 1297 
%\bibitem[1978]{bohlin}
%Bohlin, R.C., Savage, B.D., Drake, J.F. 1978, ApJ, 224, 132
\bibitem[2001]{brand}
Brand, J., Cesaroni, R., Palla, F., Molinari, S. 2001, A\&A, 370, 230
\bibitem[1998]{carey}
Carey, S.J., Clark, F.O., Egan, M.P. et al. 1998, ApJ, 508, 721
\bibitem[1999]{caselli99} 
Caselli, P., Walmsley, C.~M., Tafalla, M., Dore, L., \& Myers, P.~C.\ 1999, 
ApJ, 523, L165
\bibitem[2002a]{casellia}
Caselli, P., Walmsley C.M., Zucconi, A. et al. 2002a, ApJ, 565, 331
\bibitem[2002b]{casellib}
Caselli, P., Walmsley C.M., Zucconi, A. et al. 2002b, ApJ, 565, 344
\bibitem[2002c]{casellic}
Caselli, P., Benson, P.J., Myers, P.C., Tafalla, M. 2002c, ApJ, 572, 238
\bibitem[2005]{caselli05}
Caselli, P. 2005, proceedings of the "Cores to Clusters" meeting, Porto, Portugal, October 2004
\bibitem[2003]{cesa03}
Cesaroni, R.; Codella, C.; Furuya, R. S.; Testi, L. 2003, A\&A, 401, 227
\bibitem[2003]{collings}
Collings, M.~P., Dever, J.~W., Fraser, H.~J., McCoustra, M.~R.~S., \& 
Williams, D.~A.\ 2003, ApJ, 583, 1058 
\bibitem[2004]{crapsi04}
Crapsi, A., Caselli, P., Walmsley, C.~M., Tafalla, M., Lee, C.~W., Bourke, 
T.~L., \& Myers, P.~C.\ 2004, A\&A, 420, 957 
\bibitem[2005]{crapsi}
Crapsi, A., Caselli, P., Walmsley, C.M., et al. 2005, ApJ, 619, 379
\bibitem[2006]{daniel}
Daniel, F., Chernicharo, J., Dubernet M.L. 2006, astro-ph/0606479
\bibitem[2004]{dore}
Dore, L., Caselli, P., Beninati, S. et al. 2004, A\&A, 413, 1177
\bibitem[2001]{evans}
Evans, N.J.II, Shirley, Y.L. \& Mundy, L.G. 2001, ApJ, 557, 193
\bibitem[2003]{flower03} 
Flower, D.~R., \& Pineau des For{\^e}ts, G.\ 2003, MNRAS, 343, 390
\bibitem[2005]{flower05}
Flower, D.~R., Pineau des For{\^e}ts, Walmsley, C.M. 2005, A\&A, 436, 933
\bibitem[2004a]{fonta1}
Fontani, F., Cesaroni, R., Testi, L. et al. 2004a, A\&A, 414, 299 
\bibitem[2004b]{fonta2}
Fontani, F., Cesaroni, R., Testi, L. et al. 2004b, A\&A, 424, 179 
\bibitem[2005]{fonta05}
Fontani, F., Beltr\'an, M.T., Brand, J. et al. 2005, A\&A, 432, 921
\bibitem[1982]{frerking}
Frerking, M., Langer, L., Wilson, R. 1982, ApJ, 262, 590
\bibitem[2005]{fuller}
Fuller, G.A., Williams, S.J., Sridharan, T.K. 2005, A\&A, 442, 949
\bibitem[2002]{gerlich}
Gerlich, D., Herbst, E., Roueff, E. 2002, P\&SS, 50.1275
\bibitem[2003]{harvey}
Harvey, D.W.A., Wilner, D.J., Lada, C.J., Myers, P.C., \& Alves, J.F. 2003, 
ApJ, 598, 1112
\bibitem[1992]{hasegawa92}
Hasegawa, T.I., Herbst, E., Leung, C.M. ApJS 1992, 82, 167
\bibitem[1993]{hasegawa93}
Hasegawa, T.I., Herbst, E. 1993, MNRAS, 263, 589
\bibitem[2000]{hatchell}
Hatchell, J., Fuller, G.A., Millar, T.J., Thompson, M.A., Macdonald, G.H. 2000, A\&A, 357, 637 
\bibitem[2003]{hatchell03}
Hatchell, J. 2003, A\&A, 403, 25L
\bibitem[2000]{hofner}
Hofner, P., Wyrowski, F., Walmsley, C. M., Churchwell, E. 2000, ApJ, 536, 393
\bibitem[1998]{holland}
Holland, W.S., Cunningham, C.R., Gear, W.K. et al. 1998, SPIE, 3357, 305
\bibitem[1999]{jijina}
Jijina, J., Myers, P.C., Adams, F.C. 1999, ApJS, 125, 161
%\bibitem[1989]{jarrett}
%Jarrett, T.H., Dickman, R.L., Herbst, W. 1989, ApJ, 345, 881
\bibitem[1998]{kramer}
Kramer, C., Alves, J., Lada, C., et al. 1998, A\&A, 329, L33
\bibitem[2003]{kramer03}
Kramer, C., Richer, J., Mookerjea, B., Alves, J., Lada, C. 2003, A\&A, 399, 1073
\bibitem[1996]{kuiper}
Kuiper, T.~B.~H., Langer, W.~D., \& Velusamy, T.\ 1996, ApJ, 468, 761
\bibitem[2000]{kurtz}
Kurtz, S., Cesaroni, R., Churchwell, E., Hofner, P., Walmsley, C.M. 2000, 
Protostars and Planets IV, 299
\bibitem[1981]{kutner}
Kutner, M.L. \& Ulich, B.L. 1981, ApJ, 250, 341
\bibitem[1994]{lacy}
Lacy, J.H., Knacke, R., Geballe, T.R., Tokunaga, A.T. 1994, ApJ, 428, L69
\bibitem[2002]{loinard}
Loinard, L., Castets, A., Ceccarelli, C. et al. 2002, P\&SS, 50, 1205
\bibitem[1977]{mathis77}
Mathis, J.S., Rumpl, W., Nordsieck, K. H. 1977, ApJ, 217, 425
%\bibitem[1990]{mathis}
%Mathis, J.S. 1990, ARA\&A, 28, 37
\bibitem[1989]{mckee}
McKee, C.F. 1989, ApJ, 345, 782
\bibitem[1996]{mol96}
Molinari, S., Brand, J., Cesaroni, R., Palla, F. 1996, A\&A, 308, 573
\bibitem[1998]{mol98}
Molinari, S., Testi, L., Brand, J., Cesaroni, R., Palla, F. 1998, ApJ, 505, L39 
\bibitem[2000]{mol00}
Molinari, S., Brand, J., Cesaroni, R., Palla, F. 2000, A\&A, 355, 617
\bibitem[2002]{mol02}
Molinari, S., testi, L. Rodriguez, L.F., Zhang, Q. 2002, ApJ, 570, 758
\bibitem[2001]{mea}
Motte, F., Andr\'e P. 2001, A\&A, 365, 440
\bibitem[2005]{oberg}
\"Oberg, K.I., van Broekhuizen, F., Fraser, H.J. et al. 2005, ApJ, 621L, 33  
\bibitem[1984]{oloffson}
Oloffson, H. 1984, A\&A, 134, 36
\bibitem[2006]{parise}
Parise, B., Ceccarelli, C., Tielens, A.G.G.M. et al. 2006, A\&A, 453, 949
%\bibitem[1991]{palla}
%Palla F., Brand J., Cesaroni R., Comoretto G., Felli M. 1991, A\&A, 246, 249
\bibitem[2003]{roberts03}
Roberts, H.; Herbst, E.; Millar, T. J. 2003, ApJ, 591, 41L
\bibitem[2004]{roberts04}
Roberts, H.; Herbst, E.; Millar, T. J. 2004, A\&A, 424, 905
\bibitem[1990]{sandford}
Sandford, S.~A., \& Allamandola, L.~J.\ 1990, Icarus, 87, 188 
\bibitem[1977]{shu}
Shu, F.H. 1977, ApJ, 214, 488
\bibitem[2002]{sridharan}
Sridharan, T.K., Beuther, H., Schilke, P., Menten, K.M., Wyrowski, F. 2002, ApJ, 566, 931
\bibitem[2005]{sridharan05}
Sridharan, T.K., Beuther, H., Saito, M., Wyrowski, F., Schilke, P. 2005, A\&A, 634, L57 
\bibitem[2002]{tafalla02}
Tafalla, M., Myers, P.C., Caselli, P., Walmsley, C.M., Comito, C. 2002, ApJ, 
569, 815
\bibitem[2004]{tafalla04}
Tafalla, M., Myers, P.~C., Caselli, P., Walmsley, C.~M.\ 2004, A\&A, 416, 
191
\bibitem[2006]{tafalla06} 
Tafalla, M., Santiago, J., Myers, P.~C., Caselli, P., Walmsley, C.~M., \& 
Crapsi, A.\ 2006, A\&A, in press (astro-ph/0605513) 
\bibitem[1987]{tea}
Tielens, A. G. G. M. \& Allamandola, L. J. 1987, Physical processes in interstellar clouds Proceedings, 333
\bibitem[1990]{turner}
Turner, B.E. 1990, ApJ, 362, L29
\bibitem[2000]{vanevan}
van der Tak, F.F.S., van Dishoeck, E.F. 2000, A\&A, 358L, 79
\bibitem[2004]{viti}
Viti, S., Collings, M.~P., Dever, J.~W., McCoustra, M.~R.~S., \& Williams, 
D.~A.\ 2004, MNRAS, 354, 1141 
\bibitem[2004]{walmsley}
Walmsley, C. M.; Flower, D. R.; Pineau des Forêts, G., 2004, A\&A, 418, 1035
\bibitem[2004]{williams}
Williams, S.J., Fuller, G.A., Sridharan, T.K. 2004, A\&A, 417, 115
\bibitem[1992]{wem}
Wilson, T.L. \& Matteucci, F. 1992, A\&AR, 4, 1
\bibitem[1994]{wer}
Wilson, T.L. \& Rood, R. 1994, ARA\&A, 32, 191

\end{thebibliography}
\end{document}